\documentclass[11pt,a4paper]{article} 
\usepackage{jheppub}
\usepackage{epsfig}
\usepackage{graphicx}
\usepackage{amsmath}
\usepackage{amsfonts}
\usepackage{amssymb}
\usepackage{slashed}
\usepackage{natbib}
\usepackage{xcolor}
\usepackage{subfigure}
\usepackage{placeins} 

\def\beq{\begin{equation}}
\def\eeq{\end{equation}}
\def\beqa{\begin{eqnarray}}
\def\eeqa{\end{eqnarray}}
\newcommand{\nn}{\nonumber}

\def\eqn#1{eq.~(\ref{#1})}
\def\Eqn#1{Eq.~(\ref{#1})}
\newcommand\eqns[2]    {eqs.\,(\ref{#1}) and~(\ref{#2})}

\newcommand\eqnss[2]   {eqs.\,(\ref{#1})--(\ref{#2})}

\newcommand\refr[1]      {ref.\,\cite{#1}}

\newcommand{\angbra}[1]{\langle #1 \rangle}
\def\la{\langle}
\def\ra{\rangle}
\def\lb{[}
\def\rb{]}

\title{The Central Emission Vertex of two gravitons}

\author[1]{Damiano Barcaro,}
\author[1]{Vittorio Del Duca}

\affiliation[1]{INFN, Laboratori Nazionali di Frascati, 00044 Frascati (RM), Italy }

\emailAdd{Vittorio.DelDuca@lnf.infn.it, damianob2000@gmail.com}

\date{today}

\abstract{It has recently been shown that there exists an $s$-channel sequence of classical corrections~\cite{Rothstein:2024nlq} to the $H$ diagram computed long ago by Amati, Ciafaloni and Veneziano~\cite{Amati:1990xe}. At leading logarithmic accuracy, those corrections feature the gravity BFKL kernel as a crucial element, and may be computed through either rapidity renormalisation group equations or amplitudes built through $s$-channel unitarity cuts.
In this paper, we evaluate six-graviton amplitudes in next-to-multi-Regge kinematics, and compute for the first time the Central Emission Vertex for the emission of two gravitons, which is relevant to evaluate the corrections to the gravity BFKL kernel, and thus to go beyond the leading logarithmic accuracy.}

\begin{document}

\maketitle

\section{Motivation}

The observation of gravitational waves by the LIGO and Virgo collaborations~\cite{LIGOScientific:2016aoc,LIGOScientific:2017vwq} and the ensuing high precision measurements and cataloguing of compact binary mergers by the LIGO-Virgo-KAGRA (LVK) network~\cite{LIGOScientific:2018mvr,LIGOScientific:2020ibl,KAGRA:2021vkt} has started the age of gravitational-wave
astronomy. A much larger kinematic reach, and thus much more data, is expected by the next generation of ground-\cite{Punturo:2010zz,Reitze:2019iox} and space-based~\cite{LISA:2017pwj} gravitational-wave observatories.

In the quest to match the increasing accuracy of the experimental data, the theory community\footnote{Responding also to an invitation~\cite{Damour:2017zjx}.} has been providing more and more accurate waveform models of binary systems, based on post-Newtonian (PN)~\cite{Blanchet:2002av,Foffa:2013qca,Foffa:2019yfl,Bini:2020nsb,Bini:2020hmy},
post-Minkowskian (PM)~\cite{Neill:2013wsa,Cheung:2018wkq,Bern:2019crd,Kalin:2020fhe,Bjerrum-Bohr:2022blt,Buonanno:2022pgc} and gravitational self-force (SF)~\cite{Poisson:2011nh,Barack:2018yvs,Pound:2021qin} expansions in combination with effective field theory (EFT) methods~\cite{Goldberger:2004jt,Porto:2016pyg}, to be compared to numerical relativity (NR) simulations~\cite{Pretorius:2005gq,Campanelli:2005dd,Baker:2005vv}, and eventually to be combined with them. 

The PM expansion is strictly related to the classical part of the loop expansion of scattering amplitudes in Newton constant $G$: an $n-$loop four-point amplitude is ${\cal O}(G^{n+1})$ and contributes to the $(n+1)$PM order. Using amplitude and EFT techniques~\cite{Kalin:2020mvi,Mogull:2020sak}, complete results up to three loops (4PM)~\cite{Bern:2021dqo,Dlapa:2021npj,Bern:2021yeh,Dlapa:2021vgp,Dlapa:2022lmu,Jakobsen:2023ndj,Jakobsen:2023hig,Damgaard:2023ttc,Dlapa:2024cje}, and for the 1SF part of the four-loop (5PM) computation~\cite{Driesse:2024xad,Driesse:2024feo,Dlapa:2025biy} have been achieved.

In particular, it has been found that in the high-energy (Regge) limit~\cite{Regge:1959mz}, defined as the limit in which the squared center-of-mass energy $s$ is much larger than the squared momentum transfer $t$, 
the imaginary part of the two-loop $2 \to 2$ amplitude, ${\rm Im}[{\cal M}_4^{(2)}]$, exhibits a $\log(s/|t|)$ behaviour at high energies. That $\log(s/|t|)$ behaviour is universal, i.e. it is independent of the mass and spin of the scattering particles~\cite{DiVecchia:2020ymx}.
${\rm Im}[{\cal M}_4^{(2)}]$ is provided through a three-particle cut by the $H$ diagram computed by Amati, Ciafaloni and Veneziano (ACV)~\cite{Amati:1990xe} in multi-Regge kinematics, and is related to the radiation-reaction (RR) piece of the two-loop (3PM) scattering angle.
However, due to a crucial cancellation occurring between the conservative and RR contributions, the scattering angle at 3PM does not contain any high-energy logarithmic divergence~\cite{DiVecchia:2020ymx,Herrmann:2021tct,Jakobsen:2022fcj,Alessio:2022kwv}. 
A high-energy logarithm occurs instead in the scattering angle at 4PM~\cite{Dlapa:2022lmu}.
Finally, it has recently been shown that in the Regge limit there is an $s$-channel sequence of classical corrections of ${\cal O}\left( (G^2 s \log(s/|t|))^n \right)$ to the ACV computation~\cite{Rothstein:2024nlq}.

This raises interesting questions: going to higher PM orders in the Regge limit, do $2 \to 2$ amplitudes display universal features? Do they exhibit a pattern of high-energy logarithms? If so, how is the scattering angle affected by those features and that pattern?

In the Regge limit, any $2 \to 2$ scattering process is dominated by the exchange in the $t$ channel of the highest-spin particle~\cite{Gribov:1970ik,tHooft:1987vrq}. That entails the exchange in the $t$ channel of a vector boson carrying the $SU(N)$ Yang-Mills interaction in a non-Abelian gauge theory, and thus of a gluon in QCD, and of a graviton in a gravity theory.
Contributions that do not feature gluon or graviton exchange in the $t$ channel are power suppressed in $t/s$. 

Radiative corrections to the $2 \to 2$ scattering amplitudes display iterative patterns of the evolution in the rapidity $y\simeq \log(s/|t|)$, either if the evolution occurs in the $t$ channel or in the $s$ channel~\cite{Lipatov:1989bs}. 
Which evolution is prevailing, it depends then on the theory at hand: in QCD, the leading radiative corrections are due to a gluon ladder, associated to a gluon exchanged in the $t$ channel~\cite{Lipatov:1976zz,Kuraev:1976ge}, termed Reggeised gluon or briefly Reggeon in Regge-Gribov theory~\cite{Gribov:2003nw} or Glauber gluon\footnote{Reggeons and Glauber gluons are not equivalent~\cite{Gao:2024qsg}, since Reggeons are $s\leftrightarrow u$ crossing symmetric, while Glauber gluons are not. However, they lead to the same BFKL equation, and in this context we will not be concerned with their differences.} in soft-collinear effective theory (SCET)~\cite{Rothstein:2016bsq}. The $s$-channel evolution terms, like the eikonal phase terms or the exchange of two Reggeons, which is the backbone of the Balitsky-Fadin-Kuraev-Lipatov (BFKL) equation~\cite{Kuraev:1977fs,Balitsky:1978ic} for singlet exchange, are logarithmically suppressed. Conversely, in gravity the leading radiative corrections are due to Weinberg's soft gravitons~\cite{Weinberg:1965nx}, while the exchange of a Reggeised or Glauber graviton, with the associated graviton trajectory, is power suppressed in $t/s$~\cite{Bartels:2012ra,Melville:2013qca}. 

In the Regge limit of QCD, in addition to the virtual corrections to the $2 \to 2$ scattering amplitudes, mentioned above, which exponentiate the 
leading logarithmic term of the one-loop $2 \to 2$ scattering amplitude and form the gluon trajectory, 
process-independent features are displayed by the radiative emissions, i.e. $2 \to n$ amplitudes with $n\ge 3$, in the multi-Regge kinematics (MRK), which orders the outgoing particles in rapidity. The $2 \to n$ amplitudes in MRK iterate the emission of a final-state gluon along the $t$-channel gluon ladder, first occurring in a $2 \to 3$ amplitude. The emission of a gluon along the ladder is termed {\it Lipatov vertex} or {\it central-emission vertex} (CEV)~\cite{Lipatov:1976zz}, and forms the kernel of an iterative integral equation, the BFKL equation, which resums the radiative corrections of ${\cal O}(\alpha_s^n \log^n(s/|t|))$ to parton-parton scattering~\cite{Lipatov:1976zz,Kuraev:1976ge,Kuraev:1977fs,Balitsky:1978ic}, with $\alpha_s$ the strong coupling constant. 
In the soft limit, the Lipatov vertex yields infrared divergences, which are regulated by the gluon trajectory.

In the Regge limit of a gravity theory, it is possible to mimic the gluon ladder construction, i.e. to consider a graviton ladder exchanged in the $t$ channel, and then to construct a gravity BFKL equation~\cite{Lipatov:1982vv,Lipatov:1982it,Lipatov:1991nf}, which resums the radiative corrections of ${\cal O}(G^n \log^n(s/|t|))$ to $2 \to 2$ scattering.
Just like in QCD, its building blocks are the graviton trajectory and the graviton CEV, with the latter building up the gravity BFKL kernel.
An interesting feature is that the graviton CEV is the double copy of the gluon CEV~\cite{Lipatov:1982vv,Lipatov:1982it}.
The graviton CEV occurs also in the context of shockwave collisions~\cite{Raj:2023irr,Raj:2023iqn,Raj:2024xsi} and of Glauber SCET of gravity~\cite{Rothstein:2024nlq}.

However, in addition to being power suppressed in $t/s$, in an expansion in $\hbar$ the graviton trajectory is a quantum correction, thus it does not contribute at all to classical or Einstein's gravity~\cite{Rothstein:2024nlq}.
So is also the emission of two or more graviton CEVs in $2 \to n$ amplitudes with $n\ge 4$~\cite{Britto:2021pud}. In fact, the whole graviton ladder exchanged in the $t$ channel is a quantum correction to classical gravity, but for the emission of a single graviton CEV, which is featured in the ACV computation of the {\it H~diagram}~\cite{Amati:1990xe}, i.e. of the imaginary part of the two-loop $2 \to 2$ amplitude, computed through a three-particle cut in MRK.

In the Glauber SCET of gravity~\cite{Rothstein:2024nlq},
the graviton ladder exchanged in the $t$ channel between two Glauber gravitons leads to the BFKL equation, which is introduced through a rapidity renormalisation group equation (RRGE), whose anomalous dimension is provided by the gravity BFKL kernel. Further, the graviton ladder exchanged in the $t$ channel between three Glauber gravitons leads also to a BFKL-like equation, whose RRGE is a convolution of two gravity BFKL kernels. Such a graviton ladder is also a quantum correction, but for a term, which is a classical correction occurring at 5PM order, thus in the four-loop $2 \to 2$ amplitude, and which is a correction of ${\cal O}(G^2 s \log(s/|t|))$ to the ACV computation. In fact, the graviton ladder exchanged in the $t$ channel between $(n+2)$ Glauber gravitons features a classical term at $(2n+3)$PM order and provides a correction of ${\cal O}\left( (G^2 s \log(s/|t|))^n \right)$ to the ACV computation, effectively building up an $s$-channel sequence of classical terms, which form a multi-H generalisation of the H~diagram~\cite{Rothstein:2024nlq}\footnote{Multi-H~diagrams were also considered in ref.~\cite{Amati:2007ak,Ciafaloni:2015xsr}.}. Interestingly, the same sequence of classical terms is featured in an iteration of $s$-channel unitarity cuts, built around tree-level amplitudes in MRK, with up to $2 \to (n+3)$ particles, thus with $(n+1)$ graviton CEVs~\cite{Alessio:2025xxx}\footnote{The $2 \to (n+3)$ amplitudes are used in such a way as to never form a cut graviton loop, thus evading the veto on the classical contribution of tree amplitudes with at least six points and the emission of at least two gravitons along the ladder~\cite{Britto:2021pud}.}.

In order to go beyond the leading logarithmic accuracy
of the $s$-channel classical sequence, we take the cue from QCD, where the next-to-leading logarithmic corrections to the BFKL equation~\cite{Fadin:1998py,Ciafaloni:1998gs,Kotikov:2000pm,Kotikov:2002ab}, are obtained by considering the radiative corrections to the leading-order kernel. Those corrections are generated by the CEV for the emission of two gluons, or a $q\bar q$ pair, close in rapidity along the gluon ladder~\cite{Fadin:1989kf,DelDuca:1995ki,Fadin:1996nw,DelDuca:1996nom,DelDuca:1996km}, and 
the one-loop corrections to the gluon CEV~\cite{Fadin:1993wh,Fadin:1994fj,Fadin:1996yv,DelDuca:1998cx,Bern:1998sc}. By $\hbar$ power counting, in gravity we expect that in the MRK limit of any amplitude with $t$-channel exchange of a single Reggeised graviton, 
loop corrections to the graviton CEV are a quantum contribution, just like is the case for the graviton trajectory.
However the CEV for the emission of two gravitons, as long as it is used in such a way to never form a graviton loop, is expected to contribute classically at next-to-leading logarithmic accuracy.
Thus, in this paper we compute for the first time the CEV for the emission of two gravitons.

The paper is organised as follows: in sec.~\ref{sec:4amps}, we consider the Regge limit of four-point amplitudes with exchange of a graviton in the $t$ channel, showing that they factorise in process-dependent impact factors and a graviton propagator. Building upon the Regge factorisation of the four-point amplitudes, in sec.~\ref{sec:5amps} we consider five-point amplitudes in
MRK and use them to derive the graviton CEV~\cite{Lipatov:1982vv,Lipatov:1982it,Lipatov:1991nf} through the spinor-helicity formalism; further, we verify that six-graviton amplitudes in MRK display the emission of two graviton CEVs, thus building up the iterative structure of the graviton ladder. In sec.~\ref{sec:nmrk}, we consider six-graviton amplitudes in next-to-multi-Regge kinematics (NMRK) and compute for the first time the two-graviton CEV.
In sec.~\ref{sec:concl}, we draw our conclusions. 

The paper is furnished with several appendices: in App.~\ref{sec:appkin}, we consider the kinematic regions which are used to evaluate the amplitudes; in App.~\ref{sec:divecchia}, we show that the amplitude for four scalars and a graviton which is used in the H~diagram~\cite{Amati:1990xe} is equivalent in MRK to the 
amplitude for four massive scalars and a graviton, evaluated in ref.~\cite{Britto:2021pud}; in App.~\ref{sec:sqcev}, we square the graviton CEV in the spinor-helicity formalism, and accordingly provide a simple derivation of the gravity BFKL kernel~\cite{Lipatov:1982it}; in App.~\ref{app:hodges},
we recall Hodges' formula for the $n$-graviton MHV amplitude~\cite{Hodges:2012ym}, and use it to compute the five-graviton amplitude; in App.~\ref{app:d}, we display three equivalent representations of the six-graviton MHV amplitude, and evaluate them in the MRK and NMRK limit, thus computing the CEV for the emission of two gravitons of equal helicities; 
in App.~\ref{app:e}, we display two equivalent representations of the six-graviton NMHV amplitude, and evaluate them in the MRK and NMRK limit, thus computing the CEV for the emission of two gravitons of opposite helicities.

\section{Regge limit of four-point amplitudes}
\label{sec:4amps}

In considering the elastic scattering $p_1 p_2 \to p_3 p_4$ of four particles, we take all momenta as outgoing, $p_1 +p_2 + p_3 + p_4 =0$. The scattering is characterised by three invariants, $s=(p_1+p_2)^2$, $t=(p_2+p_3)^2$, $u=(p_1+p_3)^2$ and two masses $m_1^2$ and $m_2^2$, linked by momentum conservation,
\beq
s + t + u = 2 (m_1^2 + m_2^2) \,,
\eeq
and mass-shell conditions,
\beq
p_1^2 = p_4^2 = m_1^2\,, \qquad p_2^2 = p_3^2 = m_2^2\,.
\eeq

The Regge limit of the $2\to 2$ amplitudes,
\beq
s \gg t, m_1^2, m_2^2 \,,
\label{eq:reggelim}
\eeq
is characterised by strong orderings in the light-cone momenta of the two final-state particles,
\beq
p_3^+\gg p_4^+\,, \qquad p_3^-\ll p_4^-\,,
\label{eq:strordpm}
\eeq
where the light-cone momenta are defined in App.~\ref{sec:appa}, and where 
the second strong ordering is equivalent to the first because of the mass-shell conditions $p_3^+p_3^- = |p_{3\perp}|^2 + m_2^2$ and $p_4^+p_4^- = |p_{4\perp}|^2 + m_1^2$, 
and of transverse momentum conservation, $p_{3\perp} + p_{4\perp} = 0$.
Since for a light-cone momentum, $p^\pm = m_\perp e^{\pm y}$, where $m_\perp = \sqrt{|p_{\perp}|^2 + m^2}$ is the transverse mass and $y$ is the rapidity,
eq.~(\ref{eq:strordpm}) is equivalent to a strong ordering of the rapidities of the final-state particles. Further, for massless particles, the transverse mass is reduced to the absolute value of the transverse momentum.

Since we will retain terms in the large ratio $s/t$, \eqn{eq:reggelim} implies that the relative size of the squared momentum transfer $t$ and the square masses $m_1^2, m_2^2$ is immaterial, i.e. terms which are functions of the ratios $t/m_1^2$, $t/m_2^2$ or $m_1^2/m_2^2$ are power suppressed in $t/s$.

Further, we will deal jointly with the massive case, with $m_1^2\ne 0$ and/or $m_2^2\ne 0$, e.g. for massive scalars or particles with spin, and with the massless case, with $m_1^2= 0$ and/or $m_2^2= 0$, e.g. for gravitons, since within the confines of our approximation the massless case can be taken as a smooth limit of the massive one.

\subsection{Four-scalar amplitude}
\label{sec:4scamps}

Firstly, we consider the interaction of two spinless particles of mass $m_1$ and $m_2$, whose fields $\phi$ are minimally coupled to gravity, $\phi_1 \phi_2 \to \phi_3 \phi_4$.
The corresponding tree amplitude is~\cite{Bern:2019crd,Britto:2021pud}
\beq
{\cal M}(1_\phi,2_\phi,3_\phi,4_\phi) = - \kappa^2 \frac{(s - m_1^2 - m_2^2)^2 - 2m_1^2 m_2^2}{t} \,,
\label{eq:4scalars}
\eeq
where the coupling $\kappa$ is related to Newton constant by the relation $\kappa^2 = 8\pi G$.

In the Regge limit (\ref{eq:reggelim}), \eqn{eq:4scalars} becomes 
\beq
\lim_{s\gg t}{\cal M}(1_\phi,2_\phi,3_\phi,4_\phi) = - \kappa^2 \frac{s^2}{t} \,,
\label{eq:4scalarsreg}
\eeq
with the amplitude pictured in fig.~\ref{fig:4pt-grav}$(a)$.

\subsection{Compton scattering}
\label{sec:4scgramps}

We consider then the Compton scattering of a spinless particle of mass $m$ and a graviton, $\phi_1 h_2 \to h_3 \phi_4$. Momentum conservation implies that
\beq
s + t + u = 2 m^2 \,.
\eeq
The Compton scattering tree amplitudes are~\cite{Arkani-Hamed:2017jhn,Guevara:2018wpp,Bjerrum-Bohr:2013bxa}
\beqa
{\cal M}(1_\phi,2_h^-,3_h^+,4_\phi) &=& \kappa^2 \frac{\langle 2 1 3 ]^4}{(s-m^2) t (u-m^2)} \,, 
\label{eq:2scalars2hlead}\\
{\cal M}(1_\phi,2_h^+,3_h^+,4_\phi) &=& \kappa^2 \frac{ m^4 [2 3 ]^4}{(s-m^2) t (u-m^2)} \,,
\label{eq:2scalars2hnext}
\eeqa
where the superscript on gravitons labels the helicity\footnote{With a slight abuse of notation, helicity is labelled by the same $\pm$ sign used in light-cone momenta, but being used in a different context it should not generate confusion.}
\footnote{In our all-outgoing-momenta convention, helicity labels for incoming particles are the negative of their physical helicities.}, and where spinor products are defined in App.~\ref{sec:appa}. The amplitudes ${\cal M}(1_\phi,2_h^+,3_h^-,4_\phi)$ and ${\cal M}(1_\phi,2_h^-,3_h^-,4_\phi)$ are obtained by a parity transformation, which flips all helicities and conjugates spinors, $\langle pk \rangle \leftrightarrow [kp]$.
In the Regge limit (\ref{eq:strordpm}), taking the spinor products as in App.~\ref{sec:appb}, eqs.~(\ref{eq:2scalars2hlead}) and (\ref{eq:2scalars2hnext}) become
\beqa
\lim_{s\gg t}{\cal M}(1_\phi,2_h^-,3_h^+,4_\phi) &=& - \kappa^2 \frac{s^2}{t} \,,
\label{eq:2scalars2hleadreg}\\
\lim_{s\gg t}{\cal M}(1_\phi,2_h^+,3_h^+,4_\phi) &=& - \kappa^2 \frac{ m^4 (p_{3\perp}^\ast)^4}{s^2 t} \,,
\label{eq:2scalars2hnextreg}
\eeqa
thus, we see that the helicity-flip amplitudes ${\cal M}(1_\phi,2_h^+,3_h^+,4_\phi)$ are power suppressed. Note that by swapping graviton and scalar labels, the Compton amplitude can be written as
\beq
{\cal M}(1_h^-,2_\phi,3_\phi,4_h^+) = \kappa^2 \frac{\langle 1 2 4 ]^4}{(s-m^2) t (u-m^2)} \,, 
\label{eq:2scalars2hleadb}
\eeq
whose Regge limit is
\beq
{\cal M}(1_h^-,2_\phi,3_\phi,4_h^+) = - \kappa^2 \frac{s^2}{t} \left( \frac{p_{4\perp}^\ast}{p_{4\perp}} \right)^2 \,.
\label{eq:2scalars2hleadregb}
\eeq
The amplitudes (\ref{eq:2scalars2hleadreg}) and (\ref{eq:2scalars2hleadregb}) are pictured in 
figs.~\ref{fig:4pt-grav}$(b)$ and $(c)$.

\subsection{Four-graviton amplitude}
\label{sec:4gramps}

\begin{figure}
  \centerline{\includegraphics[width=0.85\columnwidth]{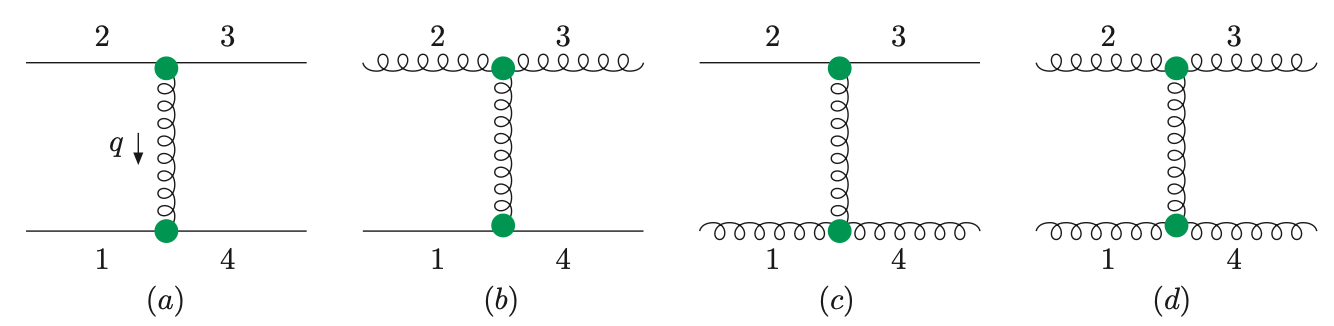} }
  \caption{Four-point amplitudes in the Regge limit. The green blobs represent  the impact factors.
  $(a)$: The four-scalar amplitude. 
  $(b)$ and $(c)$: Compton amplitudes.
  $(d)$: The four-graviton amplitude.}
\label{fig:4pt-grav}
\end{figure}

Finally, we consider the scattering of four gravitons, $h_1 h_2 \to h_3 h_4$, with momentum conservation,
\beq
s + t + u = 0\,.
\eeq
The four-graviton MHV amplitude is~\cite{Berends:1988zp},
\beq
{\cal M}(1_h^-,2_h^-,3_h^+,4_h^+) = \kappa^2 \frac{ \langle 12 \rangle^7 [12]}{\langle 13 \rangle \langle 14 \rangle \langle 23 \rangle \langle 24 \rangle \langle 34 \rangle^2} \,,
\label{eq:4h}
\eeq
where the amplitude for any other of the $\binom{4}{2} = 6$ helicity configurations of type $(-,-,+,+)$ is obtained by permuting the labels, while the amplitudes for the helicity configurations
$(-,+,+,+)$ and $(+,+,+,+)$ vanish at tree level.
In the Regge limit, eq.~(\ref{eq:4h}) becomes
\beq
\lim_{s\gg t} {\cal M}(1_h^-,2_h^-,3_h^+,4_h^+) = - \kappa^2 \frac{s^2}{t} \left( \frac{p_{4\perp}^\ast}{p_{4\perp}} \right)^2 \,,
\label{eq:4hreg}
\eeq
with the amplitude pictured in fig.~\ref{fig:4pt-grav}$(d)$.

\subsection{Regge factorisation of four-point amplitudes}
\label{sec:4regamps}

Eqs.~(\ref{eq:4scalarsreg}), (\ref{eq:2scalars2hleadreg}), (\ref{eq:2scalars2hleadregb}) and (\ref{eq:4hreg}) all have in common the exchange of a graviton in the $t$ channel, and accordingly display the factorisation of gravity amplitudes in the Regge limit, fig.~\ref{fig:4pt-grav}. They can be written as
\beq
\lim_{s\gg t} {\cal M}(1,2,3,4) = -
\left[ \kappa\, C(2,3) \right] \frac{s^2}{t}
\left[ \kappa\, C(1,4) \right] \,,
\label{eq:4regge}
\eeq
where we have introduced the scalar, $h^*\, \phi \rightarrow \phi$, and the graviton, $h^*\, h \rightarrow h$, impact factors, with $h^*$ an off-shell graviton,
\beqa
C(2_{\phi },3_{\phi }) &=& 1\,, \qquad  C(1_{\phi },4_{\phi }) = 1\,, \label{eq:gravifscal}\\
C(2^-_h,3^+_h) &=& 1 \,, \qquad 
C(1^-_h,4^+_h) = \left( \frac{p_{4\perp}^\ast}{p_{4\perp}} \right)^2 \,,
\label{eq:gravif}
\eeqa
with complex transverse coordinates $p_\perp = p_x+ i p_y$. At tree level, the graviton impact factors above are just overall phases, and they transform under parity into their complex conjugates,
\begin{equation}
[C(i_h^\nu, j_h^{\nu'})]^* = C(i_h^{-\nu}, j_h^{-\nu'})\, . 
\end{equation} 
Further, in the Regge limit, at tree level and at
leading power in $t/s$, helicity is conserved along the $s$-channel direction, or in our all-outgoing helicity convention,
\beq
C(i_h^{\nu}, j_h^{\nu'})\ \propto\ \delta^{\nu,-\nu'} .
\label{Chelcons}
\eeq
The helicity-flip impact factor $C(i_h^+, j_h^+)$ and its parity conjugate $C(i_h^-, j_h^-)$ are power suppressed in $t/s$. Thus in eq.~(\ref{eq:4hreg}) (resp. in eq.~(\ref{eq:2scalars2hleadreg})) four (resp. two) helicity configurations are leading, two for each impact factor, $h^*\, h \rightarrow h$.

Finally, note that the graviton impact factors (\ref{eq:gravif}) are the double copy of the QCD gluon impact factors, $g^*\, g \rightarrow g$~\cite{DelDuca:1995zy}.

\begin{figure}
  \centerline{\includegraphics[width=0.5\columnwidth]{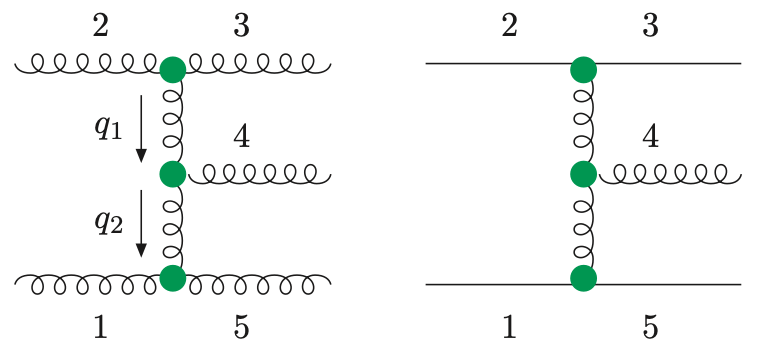} }
  \caption{Five-point amplitudes in MRK. The green blobs represent the impact factors and the graviton CEV.
  $(a)$: The five-graviton amplitude.
  $(b)$: The amplitude for four scalars and a graviton.}
\label{fig:5pt-grav}
\end{figure}

\section{Multi-Regge kinematics and the graviton CEV}
\label{sec:5amps}

Next we consider $2\to 3$ amplitudes with momenta $p_1\, p_2\to p_3\,p_4\, p_5$, and mass-shell conditions,
\beq
p_1^2 = p_5^2 = m_1^2\,, \qquad p_2^2 = p_3^2 = m_2^2\,,
\qquad p_4^2 = 0 \,,
\eeq
and let us take the outgoing momenta in multi-Regge kinematics (MRK),
\beq
p_3^+\gg p_4^+\gg p_5^+\,, \qquad |p_{3\perp}| \simeq |p_{4\perp}| \simeq |p_{5\perp}|\,,
\label{eq:mrk}
\eeq
characterised by a strong ordering in the light-cone momenta of all three final-state particles, as detailed in App.~\ref{sec:appb}, where a requirement on the transverse momenta to be all of the same size is understood\footnote{At leading accuracy in the large ratio $s/t$, the relative size of transverse mass and transverse momentum is immaterial, thus the requirement $|p_{3\perp}| \simeq |p_{4\perp}| \simeq |p_{5\perp}|$ is equivalent to $m_{3\perp} \simeq m_{4\perp} \simeq m_{5\perp}$.}.

\subsection{Regge factorisation of five-point amplitudes in MRK}
\label{sec:5regamps}

In MRK, $2\to 3$ amplitudes with exchange of a graviton in the $t$ channel, and emission of a graviton along the ensuing $t$-channel ladder, are expected to take the factorised ladder form,
\beq
\lim_{p_3^+\gg p_4^+\gg p_5^+} {\cal M}(1,2,3,4_h^{\nu_4},5) = - s^2
\left[ \kappa\, C(2,3) \right] \frac1{t_1} \left[ \kappa\, V(q_1, 4_h^{\nu_4}, q_2)\right] \frac1{t_2}
\left[ \kappa\, C(1,5) \right] \,,
\label{eq:5regge}
\eeq
with $q_1 = -(p_2+p_3)$, $q_2= p_1+p_5$, and $t_i= q_i^2\simeq - q_{i\perp} q_{i\perp}^\ast$, with $i = 1, 2$, and where the impact factors are given in \eqns{eq:gravifscal}{eq:gravif}. 
Momentum conservation requires that $p_4=q_1-q_2$.
The emission of a graviton along the graviton ladder is governed by the graviton CEV, $V(q_1, 4_h^{\nu_4}, q_2)$~\cite{Lipatov:1982vv,Lipatov:1982it,Lipatov:1991nf}.

\subsection{Five-graviton amplitude and the graviton CEV}
\label{sec:grcev}

In order to show explicitly that the five-graviton amplitude displays the ladder form (\ref{eq:5regge}), we take Hodges' formula for the $n$-graviton MHV amplitude (\ref{eq:amphodges}). For five gravitons,, one of its possible embodiments is in eq.~(\ref{eq:5happ}),
\beq
{\cal M}(1_h^-,2_h^-,3_h^+,4_h^+,5_h^+) =
\angbra{12}^8 \frac{[15][24]\angbra{14}\angbra{25}
- [14][25]\angbra{24}\angbra{15}}{\angbra{12}
    \angbra{23}\angbra{31}\angbra{34}
\angbra{45}\angbra{53}\angbra{14}\angbra{25}\angbra{24}\angbra{15}} \,.
\label{eq:5h}
\eeq
The helicity configuration of the MHV amplitude (\ref{eq:5h}) fulfills $s$-channel helicity conservation
(\ref{Chelcons}), i.e. gravitons 2 and 3 have opposite helicities, and so do gravitons 1 and 5. In MRK, eq.~(\ref{eq:5h}) becomes
\beq
\lim_{p_3^+\gg p_4^+\gg p_5^+} {\cal M}(1_h^-,2_h^-,3_h^+,4_h^+,5_h^+) = - \kappa^3 \frac{s^2}{t_1t_2} \left( \frac{p_{5\perp}^\ast}{p_{5\perp}} \right)^2 \frac{q_{1\perp}^\ast q_{2\perp}}{p_{4\perp}^2} \left( q_{1\perp}^\ast q_{2\perp} - q_{1\perp} q_{2\perp}^\ast \right)
\,,
\label{eq:5hreg}
\eeq
which displays the factorised ladder form,
\beq
\lim_{p_3^+\gg p_4^+\gg p_5^+ } {\cal M}(1_h^-,2_h^-,3_h^+,4_h^+,5_h^+) 
= - s^2
\left[ \kappa\, C(2_h^-,3_h^+) \right] \frac1{t_1} \left[ \kappa\, V(q_1, 4_h^+, q_2)\right] \frac1{t_2}
\left[ \kappa\, C(1_h^-,5_h^+) \right] \,, \label{eq:5hmhvreg}
\eeq
in agreement with eq.~(\ref{eq:5regge}), with graviton impact factors (\ref{eq:gravif}), fig.~\ref{fig:5pt-grav}$(a)$. Eq.~(\ref{eq:5hreg}) allows us to derive the CEV for a graviton of positive helicity,
\beq
V(q_1,4_h^+,q_2) = \frac{q_{1\perp}^\ast q_{2\perp}}{p_{4\perp}^2} \left( q_{1\perp}^\ast q_{2\perp} - q_{1\perp} q_{2\perp}^\ast \right)\,,
\label{eq:grlipv}
\eeq
with $p_4=q_1-q_2$, in agreement with the expression obtained in \refr{Lipatov:1991nf}. 
Note~\cite{Lipatov:1982vv,Lipatov:1982it} that \eqn{eq:grlipv} is the double copy of the QCD gluon CEV~\cite{Lipatov:1976zz,Lipatov:1991nf},
\beq
V(q_1,4_g^+,q_2) = \frac{q_{1\perp}^\ast q_{2\perp}}{p_{4\perp}}\,,
\label{eq:lipv}
\eeq
where the second term of \eqn{eq:grlipv} is there in order to avoid simultaneous singularities in the overlapping channels $s_{34}$ and $s_{45}$~\cite{Lipatov:1982vv,Lipatov:1982it}.
Under parity, eq.~(\ref{eq:grlipv}) transforms into its complex conjugate,
\beq
\left[ V(q_1,4_h^\nu,q_2) \right]^* =
V(q_1,4_h^{-\nu},q_2) \,.
\eeq

\subsection{The gravity BFKL kernel}
\label{sec:grkernel}

The gravity BFKL kernel occurs in the ultra-relativistic limit of the forward scattering amplitude~\cite{Amati:1990xe}, in
shockwave collisions~\cite{Raj:2023irr,Raj:2023iqn,Raj:2024xsi}, in
the gravity Glauber EFT~\cite{Rothstein:2024nlq}. It may be computed by taking the graviton CEV (\ref{eq:grlipv}) times itself with momentum transfer $q$ (\ref{eq:gcevplusq}), and summing over the helicity states of the graviton,
\beqa
\lefteqn{ K^h(q_1,q_2;q) } \label{eq:gravkern}\\ &=&
V(q_1,4_h^+,q_2)
 \left[V(q-q_1,-4_h^+,q-q_2) \right]^\ast + V(q_1,4_h^-,q_2) \left[V(q-q_1,-4_h^-,q-q_2) \right]^\ast \,, \nn
\eeqa
where by $-4$ we mean graviton 4 outgoing with momentum $-p_4$. In App.~\ref{sec:sqcev}, we show that
\begin{eqnarray}
K^h(q_1,q_2;q) 
&=& \left( |q|^2 - \frac{|q_1|^2 |q-q_2|^2 + |q_2|^2 |q-q_1|^2}{|q_1-q_2|^2} \right)^2 \nonumber \\ &+& 
4 \frac{|q_1|^2 |q_2|^2 |q-q_1|^2 |q-q_2|^2}{|q_1-q_2|^4}
\nonumber \\ &-& 4 \frac{ \left( q_1\cdot q_2\right)^2 |q-q_1|^2 |q-q_2|^2 }{|q_1-q_2|^4}
\nonumber \\ &-& 4\frac{ \left[ (q-q_1)\cdot (q-q_2)\right]^2 |q_1|^2 |q_2|^2 }{|q_1-q_2|^4} \,,
\label{eq:lipatovgcevmain}
\end{eqnarray}
where we omitted perp indices. \Eqn{eq:lipatovgcevmain}
agrees with the gravity BFKL kernel, eq.~(47) of \cite{Lipatov:1982it}.

\subsection{The amplitude for four massive scalars and a graviton}
\label{sec:4sc1gramps}

Next, we have taken the amplitude for four massive scalars and a graviton, as given in eq.~(4.34) of \refr{Britto:2021pud}, and verified that in the MRK limit (\ref{eq:mrk}) it takes the form,
\beq
\lim_{p_3^+\gg p_4^+\gg p_5^+}
{\cal M}(1_\phi,2_\phi,3_\phi,4_h^+,5_\phi) = - \kappa^3 \frac{s^2}{t_1t_2} \frac{q_{1\perp}^\ast q_{2\perp}}{p_{4\perp}^2} \left( q_{1\perp}^\ast q_{2\perp} - q_{1\perp} q_{2\perp}^\ast \right) \,,
\label{eq:4sc1hregge}
\eeq
i.e. the factorised ladder structure (\ref{eq:5regge})
with scalar impact factors (\ref{eq:gravifscal}) and graviton CEV (\ref{eq:grlipv}), fig.~\ref{fig:5pt-grav}$(b)$.

Likewise, in App.~\ref{sec:divecchia} we show that the amplitude for four massive scalars and a graviton, as given in eq.~(3.1) of \refr{DiVecchia:2020ymx}, reduces to \eqn{eq:4sc1hregge}, which amounts to showing the equivalence between the graviton CEV with Lorentz indices, \eqns{eq:b1}{eq:jmunu}, and the one obtained through spinor-helicity methods in \eqn{eq:grlipv}.

\subsection{Regge factorisation of six-point amplitudes in MRK}
\label{sec:6regamps}

\begin{figure}
  \centerline{\includegraphics[width=0.4\columnwidth]{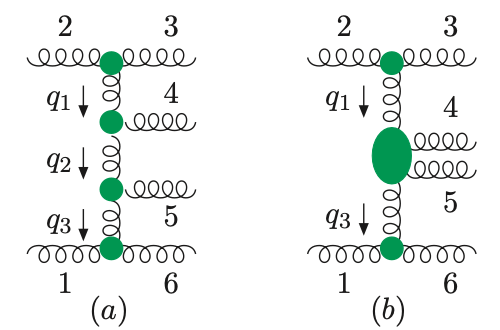} }
  \caption{Six-graviton amplitudes. $(a)$ in MRK: the green blobs represent the impact factors and the graviton CEVs. $(b)$ in NMRK: the oval green blob represents the two-graviton CEV.}
\label{fig:6pt-grav}
\end{figure}

$2\to 4$ amplitudes with momenta $p_1\, p_2\to p_3\,p_4\, p_5\, p_6$, and mass-shell conditions,
\beq
p_1^2 = p_6^2 = m_1^2\,, \qquad p_2^2 = p_3^2 = m_2^2\,,
\qquad p_4^2 = p_5^2 = 0 \,,
\label{eq:sixmass}
\eeq
and with the outgoing momenta in MRK, 
\beq
p_3^+\gg p_4^+ \gg p_5^+\gg p_6^+\,, \qquad |p_{3\perp}| \simeq |p_{4\perp}| \simeq |p_{5\perp}| \simeq |p_{6\perp}|\,,
\label{eq:mr2}
\eeq
which display the exchange of a graviton in the $t$ channel, and emission of two gravitons along the $t$-channel ladder, are expected to take the factorised ladder form, fig.~\ref{fig:6pt-grav}$(a)$,
\beqa
\lefteqn{
\lim_{p_3^+\gg p_4^+\gg p_5^+ \gg p_6^+} {\cal M}(1,2,3,4_h^\pm,5_h^\pm,6) } \nn\\
&=& - s^2
\left[ \kappa\, C(2,3) \right] \frac1{t_1} \left[ \kappa\, V(q_1, 4_h^\pm, q_2)\right] \frac1{t_2}
\left[ \kappa\, V(q_2, 5_h^\pm, q_3)\right] \frac1{t_3}
\left[ \kappa\, C(1,6) \right] \,,
\label{eq:6mrkregge}
\eeqa
with $q_1 = -(p_2+p_3)$, $q_2=q_1-p_4$, $q_3= p_1+p_6$, and $t_i= q_i^2\simeq - q_{i\perp} q_{i\perp}^\ast$, with $i = 1, 2, 3$, and impact factors given in \eqns{eq:gravifscal}{eq:gravif}. Momentum conservation requires that $p_5=q_2-q_3$. The emission of the gravitons along the graviton ladder is governed by the graviton CEV (\ref{eq:grlipv}).

For six gravitons, using Hodges' formula, \eqnss{eq:amphodges}{eq:phiij}, we display three equivalent representations of the MHV amplitude, \eqnss{eq:6gravamp1}{eq:6gravamp3}. As detailed in App.~\ref{app:dmrk}, in the MRK limit each of \eqnss{eq:6gravamp1}{eq:6gravamp3} takes the factorised ladder form (\ref{eq:6mrkregge}), with graviton
impact factors given in \eqn{eq:gravif}. Likewise, in App.~\ref{app:emrk} we report that the NMHV amplitudes (\ref{eq:nmhv6grav1}) and (\ref{eq:nmhv6grav2}) are reduced to the ladder form (\ref{eq:6mrkregge}).

\section{Next-to-multi-Regge kinematics and the two-graviton CEV}
\label{sec:nmrk}

In next-to-multi-Regge kinematics (NMRK), see App.~\ref{sec:appd},
\beq
p^+_3 \gg p^+_4 \simeq p^+_5 \gg p^+_6\,, \qquad |p_{3\perp}|
\simeq |p_{4\perp}| \simeq |p_{5\perp}| \simeq |p_{6\perp}|\, ,
\label{eq:nmreq}
\eeq
$2\to 4$ amplitudes with mass-shell condition (\ref{eq:sixmass}) and with exchange of a graviton in the $t$ channel, and emission of two gravitons along the $t$-channel ladder with no relative order in rapidity, are expected to take the factorised ladder form, fig.~\ref{fig:6pt-grav}$(b)$,
\beqa
\lefteqn{
\lim_{p_3^+\gg p_4^+\simeq p_5^+ \gg p_6^+} {\cal M}(1,2,3,4_h^{\nu_4},5_h^{\nu_5},6) } \nn\\
&=& - s^2
\left[ \kappa\, C(2,3) \right] \frac1{t_1} \left[ \kappa^2\, V(q_1, 4_h^{\nu_4}, 5_h^{\nu_5}, q_3)\right] \frac1{t_3}
\left[ \kappa\, C(1,6) \right] \,,
\label{eq:6regge}
\eeqa
with $q_1 = -(p_2+p_3)$, $q_3= p_1+p_6$, and $t_i= q_i^2\simeq - q_{i\perp} q_{i\perp}^\ast$, with $i = 1, 3$, and impact factors given in \eqns{eq:gravifscal}{eq:gravif}. Momentum conservation requires that $p_4+p_5=q_1-q_3$. 

In eq.~(\ref{eq:6regge}), the emission of two gravitons along the ladder is given by the two-graviton CEV, $V(q_1, 4_h^{\nu_4}, 5_h^{\nu_5}, q_3)$, which can be derived from six-graviton amplitudes. In particular, in the NMRK limit (\ref{eq:nmreq}), the
MHV amplitudes, \eqnss{eq:6gravamp1}{eq:6gravamp3}, are reduced to \eqn{eq:dam_NMHV}, which takes the ladder form,
\beqa
\lefteqn{ \lim_{p_3^+\gg p_4^+\simeq p_5^+ \gg p_6^+} {\cal M}(1_h^-,2_h^-,3_h^+,4_h^+,5_h^+,6_h^+) } \nn\\
&=& - s^2
\left[ \kappa\, C(2_h^-,3_h^+) \right] \frac1{t_1} \left[ \kappa^2\, V(q_1, 4_h^+, 5_h^+, q_3)\right] \frac1{t_3}
\left[ \kappa\, C(1_h^-,6_h^+) \right] \,, \label{eq:6hmhvnmrk}
\eeqa
with the CEV for two gravitons of positive helicities,
\begin{equation}
\begin{aligned}
V(q_1, 4_h^+, 5_h^+, q_3) &= \left( \sqrt{\dfrac{p_4^+}{p_5^+}} \dfrac{-p_{5\perp}}{\langle 45\rangle} \right)
\dfrac{q_{1\perp}^* q_{3\perp}}{p_{4\perp}^2p_{5\perp}^2}
\left(q_{1\perp}^*p_{4\perp} - q_{1\perp} p_{4\perp}^* \right)
\left(q_{3\perp}^*p_{5\perp} - q_{3\perp}p_{5\perp}^* \right) \\
&+(4 \leftrightarrow 5) \,.
\end{aligned}
\label{eq:2gravcev}
\end{equation}
We can also write \eqn{eq:2gravcev} as
\begin{equation}
\begin{aligned}
V(q_1, 4_h^+, 5_h^+, q_3) &= \sqrt{\dfrac{p_4^+}{p_5^+}} \dfrac{-p_{5\perp}}{\langle 45\rangle}
\dfrac{q_{1\perp}^* q_{3\perp}}{p_{4\perp}^2 p_{5\perp}^2}
\left(q_{1\perp}^*q_{2\perp} - q_{1\perp} q_{2\perp}^* \right)
\left( q_{2\perp} q_{3\perp}^* - q_{2\perp}^* q_{3\perp} \right)
\\ &+
\sqrt{\dfrac{p_5^+}{p_4^+}} \dfrac{-p_{4\perp}}{\langle 54\rangle}
\dfrac{q_{1\perp}^* q_{3\perp}}{p_{4\perp}^2 p_{5\perp}^2}
\left(q_{1\perp}^* q'_{2\perp} - q_{1\perp} {q'_{2\perp}}^{\hspace{-0.2cm}*} \right)
\left( q'_{2\perp} q_{3\perp}^* - {q'_{2\perp}}^{\hspace{-0.2cm}*} q_{3\perp} \right) \,,
\end{aligned}
\label{eq:2gravcevb}
\end{equation}
with $q_2=q_1-p_4$ and $q'_2=q_1-p_5$. In the MRK limit, \eqn{eq:2gravcevb} yields straightforwardly \eqn{eq:6hmhvregexpl}.

Likewise, in the NMRK limit (\ref{eq:nmreq}), the
NMHV amplitudes (\ref{eq:nmhv6grav1}) and (\ref{eq:nmhv6grav2}) are reduced to
\beqa
\lefteqn{ \lim_{p_3^+\gg p_4^+\simeq p_5^+ \gg p_6^+} {\cal M}(1_h^-,2_h^-,3_h^+,4_h^-,5_h^+,6_h^+) } \nn\\
&=& - s^2
\left[ \kappa\, C(2_h^-,3_h^+) \right] \frac1{t_1} \left[ \kappa^2\, V(q_1, 4_h^-, 5_h^+, q_3)\right] \frac1{t_3}
\left[ \kappa\, C(1_h^-,6_h^+) \right] \,, \label{eq:6hnmhvnmrk}
\eeqa
with the CEV for two gravitons of opposite helicity
$V(q_1, 4_h^-, 5_h^+, q_3)$ given in
eq.~(\ref{eq:nmhv2gravcev}).

Under parity, the two-graviton CEVs (\ref{eq:2gravcev}) and (\ref{eq:nmhv2gravcev}) transform into their complex conjugates,
\beq
\left[ V(q_1,4_h^{\nu_4},5_h^{\nu_5},q_2) \right]^* =
V(q_1,4_h^{-\nu_4},5_h^{-\nu_5},q_2) \,.
\eeq

\section{Conclusions}
\label{sec:concl}

In this paper, we have verified the Regge factorisation of four-, five- and six-point amplitudes which feature the exchange of a graviton in the $t$ channel, and we have re-derived the graviton CEV in the spinor-helicity formalism.

The main item of the paper, though, is the computation for the first time of the two-graviton CEV, given in eqs.~(\ref{eq:2gravcev}) and (\ref{eq:nmhv2gravcev}), through the evaluation of six-graviton amplitudes in next-to-multi-Regge kinematics. As expected, the CEV for two gravitons of positive helicities (\ref{eq:2gravcev}), which comes from MHV amplitudes, has a simple form; while the CEV for two gravitons of opposite helicities (\ref{eq:nmhv2gravcev}), which comes from NMHV amplitudes, is very involved. The two-graviton CEV will be useful in evaluating the next-to-leading-order corrections to the gravity BFKL kernel, which we leave to future work.

\section*{Acknowledgements}
We thank Cliff Cheung, Radu Roiban, Michael Ruf, Jaroslav Trnka and Raju Venugopalan for enlightening discussions. We thank Francesco Alessio, Riccardo Gonzo, Emanuele Rosi, Ira Rothstein and Michael Saavedra for the collaboration on ref.~\cite{Alessio:2025xxx}, which partly motivated the present work.
We are deeply indebted to Emmet Byrne for spotting that one of the coefficients of the two-graviton CEV of opposite helicities did not have the right scaling, which led us to find a typo in the NMHV six-graviton amplitude of ref.~\cite{Hodges:2011wm}, and 
gave us the chance to correct the two-graviton CEV of opposite helicities; we thank Emmet Byrne also for checking numerically later that the two-graviton CEV thus amended was indeed correct.

\appendix
\section{Kinematic regions}
\label{sec:appkin}

\subsection{Multi-particle kinematics}
\label{sec:appa}

We consider the scattering between two particles of  
momenta $p_1$ and $p_2$ and masses $m_1$ and $m_2$, with
production of up to four particles of momenta $p_i$, $i=3,4,5,6$, with mass-shell conditions, $p_1^2 = p_6^2 = m_1^2$ and $p_2^2 = p_3^2 = m_2^2$, and $p_4^2 = p_5^2 =0$. By convention,
we consider the scattering in the unphysical region where all momenta 
are taken as outgoing, and then we analytically continue to the
physical region where $p_1^0<0$ and $p_2^0<0$. Thus
particles are incoming or outgoing depending on the sign
of their energy. Since the helicity of a positive-energy 
(negative-energy) massless spinor has the same (opposite) sign as its
chirality, the helicities assigned to the particles 
depend on whether they are incoming or outgoing.
We label outgoing (positive-energy) particles 
with their helicity. If they are incoming the 
actual helicity and charge are reversed, e.g.~an incoming left-handed particle is labelled as an outgoing right-handed anti-particle.

Using light-cone coordinates $p^{\pm}= p^0\pm p^z $, and
complex transverse coordinates $p_{\perp} = p^x + i p^y$, with scalar product,
\begin{equation}
2 p\cdot q = p^+q^- + p^-q^+ - p_{\perp} q^*_{\perp} - p^*_{\perp} q_{\perp}\,, 
\label{eq:scalprod}
\end{equation} 
and defining the rapidity $y$ and the transverse mass $m_\perp$,
\beq
y = \frac1{2} \log\frac{p^0 + p^z}{p^0 - p^z} \,, \qquad
m_\perp = \sqrt{|p_{\perp}|^2 + m^2} \,,
\eeq
the four-momenta are
\begin{eqnarray}
p_1 &=& \left(\frac{p_1^+ + p_1^-}{2}, 0, 0,\frac{p_1^+ -p_1^-}{2} \right) 
     \equiv  \left(\frac{m_1^2}{p_1^-}, p_1^-; 0, 0\right)\, ,\nonumber \\
p_2 &=& \left(\frac{p_2^+ + p_2^-}{2}, 0, 0,\frac{p_2^+ - p_2^-}{2} \right) 
     \equiv  \left(p_2^+ , \frac{m_2^2}{p_2^+}; 0, 0 \right)\,  ,\label{in}\\
p_k &=& \left( \frac{p_k^+ + p_k^-}{2}, 
                p_k^x,
                p_k^y, 
                \frac{p_k^+ - p_k^-}{2} \right)
                \equiv
                \left( m_{k\perp} e^{y_k}, m_{k\perp} e^{-y_k}; |p_{k\perp}| \cos{\phi_k}, |p_{k\perp}| \sin{\phi_k}\right) \, ,\nonumber \\
p_i &=& \left( \frac{p_i^+ + p_i^- }{2}, 
                p_i^x,
                p_i^y, 
                \frac{p_i^+ - p_i^- }{2} \right)
                \equiv
   |p_{i\perp}|\left( e^{y_i},  e^{-y_i}; 
\cos{\phi_i}, \sin{\phi_i}\right)\, \,,\nonumber
\end{eqnarray}
with $k=3, 6$ and  $i = 4,5$. The first notation in \eqn{in} is the 
standard representation 
$p^\mu =(p^0,p^x,p^y,p^z)$ in terms of light-cone momenta;
while in the second we have the $+$ and $-$
light-cone components on the left of the semicolon, 
and the transverse components on the right, and where we have used the mass-shell conditions,
\beq
p_1^+ p_1^- = p_6^+ p_6^- - |p_{6\perp}|^2 = m_1^2 \,, \qquad p_2^+ p_2^- = p_3^+ p_3^- - |p_{3\perp}|^2 = m_2^2 \,,
\label{eq:masssh}
\eeq
such that
\beq
m_{3\perp}^2 = |p_{3\perp}|^2 + m_2^2\,, \qquad m_{6\perp}^2 = |p_{6\perp}|^2 + m_1^2 \,.
\eeq
Total momentum conservation is
\begin{eqnarray}
0 &=& \sum_{i=3}^6 p_{i\perp}\, ,\nonumber \\
p_1^+ + p_2^+ &=& -\sum_{i=3}^6 p_i^+ \, ,\label{nkin}\\ 
p_1^- + p_2^- &=& -\sum_{i=3}^6 p_i^- \, .\nonumber
\end{eqnarray}
%

\subsection{Massless momenta}
\label{sec:appaa}

The amplitudes we consider factorise into ladders made by
gravitons, bounded by either gravitons or massive scalars. For the gravitons we use the spinor-helicity formalism. Thus, in the notation of Ref.~\cite{DelDuca:1999iql}, we introduce right-handed spinor products,
$\langle k \, p \rangle = \overline{u}_{-}(k) u_{+}(p)$,
for all the momenta in \eqn{in} which are massless, 
\begin{eqnarray}\label{eq:conventions_spinors}
\langle p_i \, p_j\rangle &=& p_{i\perp}\sqrt{\frac{p_j^+}{p_i^+}} - p_{j\perp}
\sqrt{\frac{p_i^+}{p_j^+}}\, , \nonumber\\ 
\langle p_2 \, p_i\rangle &=& - i \sqrt{\frac{-p_2^+}{p_i^+}}\, p_{i\perp}\, ,\label{spro}\\ 
\langle p_i \, p_1\rangle &=&
i \sqrt{-p_1^- p_i^+}\, ,\nonumber\\ 
\langle p_2 \, p_1\rangle 
&=& -\sqrt{p_2^+p_1^-}\, ,\nonumber
\end{eqnarray}
where we have used the mass-shell condition,
\begin{equation}
|p_{i\perp}|^2 = p_i^+ p_i^-\,.
\label{eq:masssh2}
\end{equation}
Left-handed spinor products, $[k \, p] = \overline{u}_{+}(k) u_{-}(p)$, are given by complex conjugation,
\beq
[k \, p] = {\rm sign}(k^0 p^0) \langle p \, k\rangle^\ast\,.
\eeq
Spinor products are antisymmetric,
\beq
\langle k\, p\rangle = - \langle p \, k\rangle\,,\qquad [k \, p] = -[p \, k]\,.
\eeq
We also use the currents, $[p| \gamma^{\mu} |k\rangle$ and $\langle p|\gamma^{\mu}|k]$,
which are related by charge conjugation,
\begin{equation}
[p| \gamma^{\mu} |k\rangle=\langle k|\gamma^{\mu}|p]\,,
\end{equation}
and complex conjugation,
\beq
[p| \gamma^{\mu} |k\rangle^{*}  = {\rm sign}(k^0 p^0) [k| \gamma^{\mu} |p\rangle \,.
\eeq
Through the Fierz rearrangement,
\begin{equation}
 \langle k|\gamma^{\mu}|p]\langle v|\gamma^{\mu}|q] = 2 \langle k \, v\rangle [q \, p]\,,
\end{equation}
and the Gordon identity,
\begin{equation}
 [p| \gamma^{\mu} |p\rangle= \langle p|\gamma^{\mu}|p]= 2 p^{\mu}\,,
\end{equation}
we obtain that
\begin{eqnarray}
&& \langle k|\slashed{q}|p] = \langle k \, q\rangle [q \, p]\,, \nonumber\\ 
&&  [p| \slashed{q} |k\rangle = [p \, q] \langle q \, k\rangle \,.
\end{eqnarray}

Further, for a gluon of positive helicity, we choose the polarisation vector as
\beq
\epsilon_\mu^+(p,k) = \frac{[p|\gamma_\mu |k\rangle}{\sqrt{2} \langle k p\rangle} \,,
\label{eq:gravplupol}
\eeq
Later, we will use it through the double copy to define the polarisation vector of a graviton.

\subsection{Multi-Regge kinematics}
\label{sec:appb}

Extending the Regge limit (\ref{eq:reggelim}) of the $2\to 2$ amplitudes, to multi-Regge kinematics, we require that the light-cone momenta of the particles are strongly ordered and have comparable transverse momentum. In the six-point case, we take
\begin{align}
\begin{split}
&p^+_3 \gg p^+_4 \gg p^+_5 \gg p^+_6\,, \quad\qquad p^-_3 \ll p^-_4 \ll p^-_5 \ll p^-_6\,,\\
&|p_{3\perp}| \simeq |p_{4\perp}| \simeq|p_{5\perp}| \simeq|p_{6\perp}|\,,
\label{eq:mrkapp}
\end{split}
\end{align}
where, as we mentioned after eq.~(\ref{eq:mrk}), the relative size of transverse mass and transverse momentum is an effect which is suppressed in $t/s$.

The strong ordering of light-cone momenta in \eqn{eq:mrkapp} is equivalent to requiring a strong-ordering on the rapidities,
\begin{equation}
y_3 \gg y_4 \gg y_5 \gg y_6\,.
\label{MRK_hierarchy}
\end{equation}
Momentum conservation in \eqn{nkin} then becomes
\begin{eqnarray}
0 &=& \sum_{i=3}^6 p_{i\perp}\, ,\nonumber \\
p_2^+ &\simeq& -p_3^+\, ,\label{mrkin}\\ 
p_1^- &\simeq& -p_6^-\, .\nonumber
\end{eqnarray}
The Mandelstam invariants are reduced to
\begin{eqnarray}
s \simeq 2 p_1\cdot p_2 &\simeq& p_3^+ p_6^-\,, \nonumber\\ 
s_{2i} \simeq2 p_2\cdot p_i &\simeq& - p_3^+ p_i^- \,, \qquad i\ne 3\label{mrinv}\\ 
s_{1i} \simeq 2 p_1\cdot p_i &\simeq& - p_i^+ p_6^-\,, \qquad i\ne 6\nonumber\\ 
s_{ij} \simeq 2 p_i\cdot p_j &\simeq& p_i^+ p_j^-\,, \nonumber
\end{eqnarray}
with $3\le i, j \le 6$, and with $p^+_i > p^+_j$, for $i<j$. In \eqn{mrinv}, we have neglected the mass terms, which are subleading. We deal now in detail with the Mandelstam invariants for which masses cannot be neglected,
\beq
\begin{aligned}
s_{23} &= (p_2+p_3)^2 = p_2^+p_3^- + p_2^-p_3^+ +2m_2^2 \\
&\simeq -p_3^+p_3^- - \frac{m_2^2}{p_3^+} p_3^+ + 2m_2^2\\
&\simeq - |p_{3\perp}|^2 \,,
\end{aligned}
\label{eq:s23a}
\eeq
where we have used momentum conservation (\ref{mrkin}) and the mass-shell condition (\ref{eq:masssh}). It is instructive to repeat the computation by using the scalar product (\ref{eq:scalprod}) directly on the sum of momenta $p_2+p_3$,
\beq
\begin{aligned}
s_{23} &= (p_2^+ +p_3^+)(p_2^-+p_3^-) - |p_{3\perp}|^2 \\
&\simeq -p_4^+ \left( - \frac{m_2^2}{p_3^+} +p_3^- \right) - |p_{3\perp}|^2 \\
&\simeq - |p_{3\perp}|^2 \,,
\end{aligned}
\label{eq:s23b}
\eeq
where we have used momentum conservation (\ref{nkin}) and dropped subleading terms. Likewise, one shows that $s_{16} \simeq - |p_{6\perp}|^2$. 

For the three-particle invariants where masses cannot be neglected, we have
\beq
\begin{aligned}
s_{234} &= (p_2^+ +p_3^+ +p_4^+)(p_2^-+p_3^- +p_4^-) - |p_{3\perp}+p_{4\perp}|^2 \\
&\simeq -p_5^+ p_4^- - |p_{3\perp}+p_{4\perp}|^2 \\
&\simeq - |p_{3\perp}+p_{4\perp}|^2 \,.
\end{aligned}
\eeq

In fact, introducing a parameter $\lambda \ll 1$, the hierarchy in eq.~(\ref{eq:mrkapp}) above is equivalent to the rescaling,
\beq
\begin{aligned}
&s_{12} = {\cal O}(\lambda^0) \,, \qquad
\{s_{123}, s_{345} \} = {\cal O}(\lambda)\,, \\
& \{s_{34}, s_{45}, s_{56}\} = {\cal O}(\lambda^2)\,, \qquad 
\{ s_{23}, s_{61}, s_{234}, m_1^2, m_2^2 \} = {\cal O}(\lambda^3)\,.
\end{aligned}
\eeq

Note that the mass-shell condition in \eqn{eq:masssh2} implies that
\beq
s\, |p_{4\perp}|^2 |p_{5\perp}|^2 \simeq s_{34} s_{45} s_{56}\,,
\eeq
which is an example of the general multi-Regge constraint,
\beq
s_{ij}\, \prod_{k=i+1}^{j-1} |p_{k\perp}|^2 = \prod_{k=i}^{j-1} s_{k,k+1}\,.
\eeq

For massless momenta, the spinor products in \eqn{spro} become,
\begin{eqnarray}
\langle p_i \, p_j\rangle &\simeq& -\sqrt{\frac{p_i^+}{p_j^+}}\,
p_{j\perp}\, \qquad {\rm for}\, y_i>y_j \;, \nonumber\\
\langle p_2 \,  p_i\rangle &\simeq& - i\sqrt{\frac{p_3^+}{p_i^+}}\,
p_{i\perp}\, ,\label{mrpro}\\ 
\langle p_i \, p_1\rangle 
&\simeq& i\sqrt{p_i^+ p_6^-}\, ,\nonumber\\ 
\langle p_2 \,  p_1\rangle &\simeq& -\sqrt{p_3^+ p_6^-}\, ,\nonumber
\end{eqnarray}
with $3\le i \le 6$.

\subsection{Next-to-multi-Regge kinematics}
\label{sec:appd}

We now consider the production of four particles of momenta $p_i$, with $3\le i \le 6$,
with gravitons 4 and 5 in the central region along the graviton ladder,
\begin{align}
\label{nmrapp}
\begin{split}
& p^+_3 \gg p^+_4 \simeq p^+_5 \gg p^+_6\,, \quad\qquad p^-_3 \ll p^-_4 \simeq p^-_5 \ll p^-_6\,
\,, \\ 
&|p_{3\perp}|
\simeq |p_{4\perp}| \simeq |p_{5\perp}| \simeq |p_{6\perp}|\, ,
\end{split}
\end{align}
and hence,
\begin{equation}
y_3 \gg y_4 \simeq y_5\gg y_6.
\label{NMRK_hierarchy}
\end{equation}

The leading contributions to momentum conservation are the same as in \eqn{mrkin}.
The Mandelstam invariants become
\begin{eqnarray}
s \simeq 2 p_1\cdot p_2 &\simeq& p_3^+ p_6^-\,, \nonumber\\ 
s_{2i} \simeq2 p_2\cdot p_i &\simeq& - p_3^+ p_i^- \,, \qquad i\ne 3 \,, \nonumber\\ 
s_{1i} \simeq 2 p_1\cdot p_i &\simeq& - p_i^+ p_6^-\,, \qquad i\ne 6 \,, \label{nmrkinv}\\  
s_{jk} \simeq 2 p_j\cdot p_k &\simeq& p_j^+ p_k^-\,, \qquad j\ne4\,\,\, {\rm and} \,\,\, k\ne 5\,,\nonumber \\ 
s_{45} \simeq 2 p_4\cdot p_5 &=& p_4^+ p_5^- + p_4^- p_5^+ - p_{4\perp}^\ast p_{5\perp} - p_{5\perp}^\ast p_{4\perp} \nonumber\,, 
\end{eqnarray}
with $3\le i, j, k \le 6$, with $p^+_j > p^+_k$ for $j<k$. The invariants $s_{23}$ and $s_{16}$ are the same as in \eqn{eq:s23b}.

The relevant three-particle Mandelstam invariants are
\begin{align}
\begin{split}
s_{123} &\simeq (p_4^+ + p_5^+) p_6^- \,, \\
s_{234} &\simeq - |p_{3\perp} + p_{4\perp}|^2 - p_4^- p_5^+ \,, \\ 
s_{235} &\simeq - |p_{3\perp} + p_{5\perp}|^2 - p_4^+ p_5^- \,, \\ 
s_{345} &\simeq p_3^+ (p_4^- + p_5^-) \,,
\label{eq:s3nmrk}
\end{split}
\end{align}
where in $s_{234}$ masses have been properly taken into account.

Introducing a parameter $\lambda \ll 1$, the hierarchy in eq.~(\ref{nmrapp}) above is equivalent to the rescaling,
\beq
\begin{aligned}
&s_{12} = {\cal O}(\lambda^0) \,, \qquad
\{s_{34}, s_{56}, s_{123}, s_{345} \} = {\cal O}(\lambda)\,, \\ & \{ s_{23}, s_{45}, s_{61}, s_{234}, s_{235}, m_1^2, m_2^2 \} = {\cal O}(\lambda^2)\,.
\end{aligned}
\eeq

Note that the mass-shell condition in \eqn{eq:masssh} and the Mandelstam invariants in \eqn{nmrkinv} imply that
\begin{align}
\begin{split}
s &\simeq \frac{s_{34}\, s_{56} }{p_4^- p_5^+ } \\ &= \frac{s_{34}\, s_{56}\, p_4^+ p_5^- }{|p_{4\perp}|^2 |p_{5\perp}|^2}\, . \\
\end{split}
\end{align}

For massless momenta, the spinor products in \eqn{spro} become
\begin{eqnarray}
\langle p_2 \,  p_1\rangle &\simeq& - \sqrt{p_3^+ p_6^-}\,,\nonumber\\ 
\langle p_2 \, p_k\rangle &=& -i \sqrt{\frac{-p_2^+}{p_k^+}}\, p_{k\perp}
\simeq -i \sqrt{\frac{p_3^+}{p_k^+}} p_{k\perp}\, , \nonumber\\
\langle p_k \, p_1\rangle &=& i \sqrt{-p_1^- p_k^+}\, 
\simeq i \sqrt{p_k^+ p_6^-}\, ,\label{cnrpro}\\
\langle p_j \, p_k\rangle &=& p_{j\perp}\sqrt{\frac{p_k^+}{p_j^+}} - p_{k\perp}
\sqrt{\frac{p_j^+}{p_k^+}} \simeq - p_{k\perp}\, \sqrt{\frac{p_j^+}{p_k^+}}\, , \qquad j\ne4\,\,\, {\rm and} \,\,\, k\ne 5 \nonumber \\
\langle p_4 \, p_5\rangle &=& p_{4\perp}\sqrt{\frac{p_5^+}{p_4^+}} - 
p_{5\perp}\sqrt{\frac{p_4^+}{p_5^+}}\, ,\nonumber
\end{eqnarray}
with $3\le j, k \le 6$, with $p^+_j > p^+_k$. 

\section{The amplitude of four massive scalars and a graviton}
\label{sec:divecchia}

Consider the amplitude for four massive scalars and a graviton, as the field theory limit of the five-point amplitude in a bosonic string theory, eq.(3.1) of \refr{DiVecchia:2020ymx}\footnote{In \refr{DiVecchia:2020ymx} it is shown that in the Regge limit it is possible to write it as eq.(3.7) of \refr{Amati:1990xe}.}.
Using the conventions discussed in App.~\ref{sec:appkin} and the kinematic configuration shown in Fig.~\ref{fig:5pt-grav}$(a)$, the amplitude can be written as follows,
\beq
\lim_{p_3^+\gg p_4^+\gg p_5^+}
{\cal M}(1_\phi,2_\phi,3_\phi,4_h^{\mu\nu},5_\phi)
\simeq -\frac{\kappa^3 s^2}{2q_1^2q_2^2}\bigg[J^{\mu}(1,2)J^{\nu}(1,2)-q_1^2q_2^2j^{\mu}(1,2)j^{\nu}(1,2)\bigg] \,,
\label{eq:b1}
\eeq
with Lorentz indices $\mu$ and $\nu$ for the graviton, and with currents,
\beq
\begin{aligned}
&J^{\mu}(1,2)=-(q_1^{\mu}+q_2^{\mu})+q_1^2\frac{p_2^{\mu}}{p_2\cdot p_4}-q_2^2\frac{p_1^{\mu}}{p_1\cdot p_4}-\frac{s_1s_2}{s}\bigg(\frac{p_1^{\mu}}{p_1\cdot p_4}-\frac{p_2^{\mu}}{p_2\cdot p_4}\bigg) \,,\\
&j^{\mu}(1,2)=\frac{p_1^{\mu}}{p_1\cdot p_4}-\frac{p_2^{\mu}}{p_2\cdot p_4} \,.
\end{aligned}
\label{eq:jmunu}
\eeq
Note that we are taking the kinematic configuration of \refr{Amati:1990xe}, where we flip $q_1\to -q_2$, $q_2\to q_1$, and “mirror" the current associated to the incoming particles, which amounts to taking a global minus sign. 
We now proceed with the contraction with the polarization tensor of the external graviton. We consider the graviton with positive helicity, 
\begin{equation}
\epsilon_{\mu\nu}^{++}(p,k)= \epsilon_\mu^+(p,k)\epsilon_\nu^+(p,k) \,,
\end{equation}
where the amplitude with the negative helicity graviton is obtained by taking the complex conjugate of the one with positive helicity.
As we are in the Regge limit, the external massive particles behave as massless, up to $\mathcal{O}(m/E)$ corrections. Thus, it is particularly convenient to fix the reference vector of the graviton polarisation tensor to be along the light-cone direction of either $p_1$ or $p_2$, as they can be regarded as massless particles at leading order. Without loss of generality, using the gluon polarisation (\ref{eq:gravplupol}), we can therefore take the polarisation tensor of the graviton as
\begin{equation}
\epsilon_{\mu\nu}^{++}(4,1)=\frac{[p_4\gamma_\mu p_1\rangle}{\sqrt{2} \langle p_1 p_4\rangle}\frac{[p_4\gamma_\nu p_1\rangle}{\sqrt{2} \langle p_1 p_4\rangle} \,.
\label{eq:gravpluspol}
\end{equation}
Next, we contract the polarization tensor (\ref{eq:gravpluspol}) with amplitude (\ref{eq:b1}), omitting for brevity the limit sign,
\beq
\begin{aligned}
\label{eq:dam_amp}
& {\cal M}(1_\phi,2_\phi,3_\phi,4_h^{\mu\nu},5_\phi)\
\epsilon_{\mu\nu}^{++}(4,1) = 
{\cal M}(1_\phi,2_\phi,3_\phi,4_h^+,5_\phi) \\
& \qquad \simeq -\frac{\kappa^3 s^2}{2q_1^2q_2^2}
\bigg[\big(J^{\mu}(1,2)\epsilon_\mu^+(4,1) \big)^2-q_1^2q_2^2\big(j^{\mu}(1,2)\epsilon_\mu^+(4,1)\big)^2\bigg] \,,
\end{aligned}
\eeq
and study the scalar products individually.
Firstly, we consider the product,
\beq
J^\mu(1,2)\,\epsilon^+_\mu(4,1)=-2 q_2\cdot\epsilon^+(4,1) + q_1^2 \frac{p_2\cdot\epsilon^+(4,1)}{p_2\cdot p_4}+ \frac{s_1s_2}{s}\frac{p_2\cdot\epsilon^+(4,1)}{p_2\cdot p_4}\,,
\label{eq:b6}
\eeq
where we have used momentum conservation to write $q_1+q_2=2q_2+p_4$, and the identities $p\cdot\epsilon^+(p,k)=k\cdot\epsilon^+(p,k)=0$.
Since $q_2^2=-q_{2\perp}^2\neq0$, in order to write the momentum $q_2$ in spinor-helicity variables we perform a suitable shift by a massless momentum, by defining
%
\begin{align}
\bar{q}_2\equiv q_2+\alpha=\Big(-p_4^+,-\dfrac{q_{2\perp}q_{2\perp}^*}{p_4^+},q_{2\perp}\Big), \qquad  \alpha=\Big(0,-\dfrac{q_{2\perp}q_{2\perp}^*}{p_4^+},0\Big)\,,
\end{align}
i.e. we have defined the momentum $\bar{q}_2$ in such a way that it is massless by construction. Furthermore, this is also the case for the momentum $\alpha$. Finally,
note that
\begin{align}
q_2^2=(\bar{q}_2-\alpha)^2=-2\bar{q}_2\cdot\alpha=-\bar{q}_2^+\alpha^-=-q_{2\perp}q_{2\perp}^*\,,
\end{align}
which is what we would expect.
Henceforth, whenever we want to express $q_2$ in spinor-helicity variables, we may decompose it in the sum of two massless momenta and apply the usual definitions of massless spinors presented in App.~\ref{sec:appkin}. Thus
\begin{align}
q_2\cdot\epsilon^+(4,1) = (\bar{q}_2 -\alpha) \cdot\epsilon^+(4,1)
=-\dfrac{q_{1\perp}^*}{\sqrt{2}}=-\dfrac{p_{5\perp}^*}{\sqrt{2}} \,.
\end{align}
For the remaining spinor products in \eqn{eq:b6}, it is straightforward to see that
\begin{align}
\dfrac{p_2\cdot\epsilon^+(1,4)}{p_2\cdot p_4}=\dfrac{\sqrt{2}}{p_{4\perp}} \,.
\end{align}
We now recall that $q_1^2=-q_{1\perp}q_{1\perp}^*=-p_{3\perp}p_{3\perp}^*$ and $s_1s_2/s=p_{4\perp}p_{4\perp}^*$. 
Altogether we get
\beq
\begin{aligned}
J^\mu(1,2)\,\epsilon^+_\mu(4,1) &= \dfrac{\sqrt{2}}{p_{4\perp}}\left(
p_{5\perp}^*p_{4\perp}-p_{3\perp}p_{3\perp}^*+p_{4\perp}p_{4\perp}^*
\right) \\ &=
\dfrac{\sqrt{2}}{p_{4\perp}}p_{3\perp}^*p_{5\perp} \,.
\end{aligned}
\eeq
Let us now consider the product $j\cdot\epsilon^+(4,1)$ in \eqn{eq:dam_amp}, 
\begin{equation}
j^{\mu}(1,2)\epsilon^+_\mu(4,1)=-\frac{p_2\cdot\epsilon^+(4,1)}{p_2\cdot p_4}=-\dfrac{\sqrt{2}}{p_{4\perp}} \,.
\end{equation}
Finally, substituting in \eqn{eq:dam_amp}, we find as expected,
\beq
\begin{aligned}
\lim_{p_3^+\gg p_4^+\gg p_5^+}
{\cal M}(1_\phi,2_\phi,3_\phi,4_h^+,5_\phi) 
&\simeq -\kappa^3\frac{s^2}{t_1 t_2}\dfrac{p_{3\perp}^*p_{5\perp}}{p_{4\perp}^2}\left(p_{3\perp}^*p_{5\perp}-p_{3\perp}p_{5\perp}^*\right) \\ &=
-\kappa^3\frac{s^2}{t_1 t_2}\dfrac{q_{1\perp}^*q_{2\perp}}{p_{4\perp}^2}\left(q_{1\perp}^*q_{2\perp}-q_{1\perp}q_{2\perp}^*\right) \,,
\end{aligned}
\eeq
where we have defined the Mandelstam variables $t_i=q_i^2$, and used momentum conservation to replace all the external momenta $p$ with the internal ones $q$. 

Therefore, in MRK the amplitude for four massive scalars and a graviton (\ref{eq:b1}), whose graviton CEV is given by the currents in \eqns{eq:b1}{eq:jmunu}, reduces to \eqn{eq:4sc1hregge} for a graviton of positive helicity, thus showing the equivalence between the graviton CEV 
in \eqns{eq:b1}{eq:jmunu} with the one in \eqn{eq:grlipv}.

\section{Squaring the CEV}
\label{sec:sqcev}

\subsection{Yang-Mills Theory}
\label{sec:sqcevym}

In Yang-Mills theory, the gluon central emission vertex (CEV) for a gluon of positive helicity is given in \eqn{eq:lipv}, 
\begin{equation}
V(q_1,4_g^+,q_2) = \frac{q_{1\perp}^\ast q_{2\perp}}{p_{4\perp}}\,,
\label{eq:cevplus}
\end{equation}
with $p_4=q_1-q_2$. The CEV for a gluon with negative helicity is obtained by complex conjugating eq.~(\ref{eq:cevplus}).
Then squaring and summing over the two helicity states, we obtain
\begin{eqnarray}
K^g(q_1,q_2) &\equiv&
V(q_1,4_g^+,q_2) \left[V(q_1,4_g^+,q_2) \right]^\ast + V(q_1,4_g^-,q_2) \left[V(q_1,4_g^-,q_2) \right]^\ast 
\nonumber \\ &=&
2\frac{|q_{1\perp}|^2 |q_{2\perp}|^2}{|q_{1\perp}-q_{2\perp}|^2} \,.
\end{eqnarray}

In the case of the imaginary part of the forward amplitude, we need the CEV with momentum transfer $q$ with $t = q^2 = - q_\perp q_\perp^\ast$,
\begin{equation}
V(q-q_1,-4_g^+,q-q_2) = \frac{(q-q_1)_\perp^\ast (q-q_2)_\perp}{-p_{4\perp}} \,,
\label{eq:cevplusq}
\end{equation}
where by $-4$ we mean gluon 4 outgoing with momentum $-p_4=(q-q_1)-(q-q_2)$. 
Then we can define the sum over the helicity states of the gluon CEV,
\begin{eqnarray}
\lefteqn{ K^g(q_1,q_2;q) } \nonumber \\ &\equiv&
V(q_1,4_g^+,q_2) \left[V(q-q_1,-4_g^+,q-q_2) \right]^\ast + V(q_1,4_g^-,q_2) \left[V(q-q_1,-4_g^-,q-q_2) \right]^\ast 
\nonumber \\ &=&
-\frac{q_{1\perp}^\ast q_{2\perp} (q-q_1)_\perp (q-q_2)_\perp^\ast}{|q_{1\perp}-q_{2\perp}|^2} + c.c. \,,
\label{eq:cevqsum}
\end{eqnarray}
Omitting for the sake of brevity the perp indices, 
the numerator of eq.~(\ref{eq:cevqsum}) is
\begin{eqnarray}
N^g &=& q_1^\ast q_2 (q-q_1) (q-q_2)^\ast + q_1 q_2^\ast (q-q_1)^\ast (q-q_2) \nonumber \\ &=&
|q_1 (q-q_1)^\ast + q_2 (q-q_2)^\ast|^2 - |q_1|^2 |q-q_1|^2 - |q_2|^2 |q-q_2|^2 \,.
\label{eq:cevplusq2}
\end{eqnarray}
After some algebra, this can also be written as
\begin{equation}
N^g = - |q|^2 |q_1-q_2|^2 + |q_1|^2 |q-q_2|^2 + |q_2|^2 |q-q_1|^2 \,,
\label{eq:nym}
\end{equation}
such that eq.~(\ref{eq:cevqsum}) becomes
\begin{equation}
K^g(q_1,q_2;q) =
|q|^2 - \frac{|q_1|^2 |q-q_2|^2 + |q_2|^2 |q-q_1|^2}{|q_1-q_2|^2} \,,
\end{equation}
i.e. the kernel of the BFKL equation at $t\ne 0$~\cite{Lipatov:1976zz,Kuraev:1976ge,Kuraev:1977fs,Balitsky:1978ic}.

\subsection{Gravity}
\label{sec:sqcevgr}

In gravity, from eq.~(\ref{eq:grlipv}) the CEV for a graviton of positive helicity is
\begin{equation}
V(q_1,4_h^+,q_2) = \frac{q_{1\perp}^\ast q_{2\perp}}{p_{4\perp}^2} \left( q_{1\perp}^\ast q_{2\perp} - q_{1\perp} q_{2\perp}^\ast \right)\,,
\label{eq:gcevplus}
\end{equation}
and $p_4=q_1-q_2$. The CEV for a graviton of negative helicity is obtained by taking the complex conjugate,
\begin{equation}
V(q_1,4_h^-,q_2) = \left[ V(q_1,4_h^+,q_2)\right]^\ast \,.
\end{equation}
The CEV with momentum transfer $q$ is
\begin{equation}
V(q-q_1,-4_h^+,q-q_2) = \frac{(q-q_1)_\perp^\ast (q-q_2)_\perp}{p_{4\perp}^2} \left( (q-q_1)_\perp^\ast (q-q_2)_\perp - (q-q_1)_\perp (q-q_2)_\perp^\ast \right)
\,,
\label{eq:gcevplusq}
\end{equation}
where by $-4$ we mean graviton 4 outgoing with momentum $-p_4=(q-q_1)-(q-q_2)$. 
Omitting again the perp indices, the sum over the helicity states of the graviton CEV is,
\beq
\begin{aligned}
& K^h(q_1,q_2;q) \label{eq:cevqsumgrav}\\ & \equiv
V(q_1,4_h^+,q_2)
 \left[V(q-q_1,-4_h^+,q-q_2) \right]^\ast + V(q_1,4_h^-,q_2) \left[V(q-q_1,-4_h^-,q-q_2) \right]^\ast
\\ & =
\frac{q_1^\ast q_2}{p_{4\perp}^2} \left( q_1^\ast q_2 - q_1 q_2^\ast\right) \frac{(q-q_1) (q-q_2)^\ast}{(p_{4\perp}^\ast)^2}
\left( (q-q_1) (q-q_2)^\ast - (q-q_1)^\ast (q-q_2)\right)
+ c.c. \,,
\end{aligned}
\eeq
The numerator of eq.~(\ref{eq:cevqsum}) yields four terms, plus their complex conjugates.
It is straightforward to show that the square of the Yang-Mills numerator (\ref{eq:nym}) equals the first and the last terms of eq.~(\ref{eq:cevqsum}), plus their complex conjugates,
\begin{equation}
(N^g)^2 = \left( {q_1^\ast}^2 q_2^2 (q-q_1)^2 {(q-q_2)^\ast}^2 + c.c. \right) + 2 |q_1|^2 |q_2|^2 |q-q_1|^2 |q-q_2|^2 \,.
\label{eq:doubcopy}
\end{equation}
Including the second and third terms of eq.~(\ref{eq:cevqsum}), plus their complex conjugates, and using the shorthands,
\begin{eqnarray}
&& 2 q_1\cdot q_2 = q_1^\ast q_2 + q_1 q_2^\ast \nonumber \\
&& 2 (q-q_1)\cdot (q-q_2) = (q-q_1) (q-q_2)^\ast + (q-q_1)^\ast (q-q_2) \,,
\end{eqnarray}
we can write the sum over the helicity states of the graviton CEV (\ref{eq:cevqsumgrav}) as
\begin{eqnarray}
K^h(q_1,q_2;q) 
&=& \left( |q|^2 - \frac{|q_1|^2 |q-q_2|^2 + |q_2|^2 |q-q_1|^2}{|q_1-q_2|^2} \right)^2 \nonumber \\ &+& 
4 \frac{|q_1|^2 |q_2|^2 |q-q_1|^2 |q-q_2|^2}{|q_1-q_2|^4}
\nonumber \\ &-& 4 \frac{ \left( q_1\cdot q_2\right)^2 |q-q_1|^2 |q-q_2|^2 }{|q_1-q_2|^4}
\nonumber \\ &-& 4\frac{ \left[ (q-q_1)\cdot (q-q_2)\right]^2 |q_1|^2 |q_2|^2 }{|q_1-q_2|^4} \,,
\label{eq:lipatovgcev}
\end{eqnarray}
which agrees with the gravity BFKL kernel, eq.~(47) of \cite{Lipatov:1982it}.

\section{Graviton MHV amplitudes}
\label{app:hodges}

Hodges' formula for the $n$-graviton MHV amplitude $(-,-,+,+,\ldots,+)$~\cite{Hodges:2012ym} is
\begin{eqnarray}
    {\cal M}_n(1_h^{\nu_1},2_h^{\nu_2},\ldots,n_h^{\nu_n}) = \langle ij \rangle^8 M_n(1_h,2_h,\ldots,n_h) \,,
    \label{eq:amphodges}
\end{eqnarray}
where $i$ and $j$ are the negative-helicity gravitons and the reduced amplitude $M_n$ is helicity independent,
\begin{eqnarray}
    M_n(1_h,2_h,\ldots,n_h) = (-1)^{n+1} \text{sgn}(ijk)~\text{sgn}(rst) c_{ijk} c_{rst} |{\phi}|^{ijk}_{rst}  \,,
    \label{eq:redamphodges}
\end{eqnarray}
where $\text{sgn}(ijk) \equiv \text{sgn}(\sigma(i,j,k,1,2,\ldots,\slashed{i}, \slashed{j}, \slashed{k}, \ldots, n))$ is the signature of the permutation which
moves $i,j,k$ up front in the sequence, and 
\beq
c_{ijk} = \frac1{\langle ij \rangle \langle jk \rangle \langle ki \rangle } \,,
\eeq
and $|{\phi}|^{ijk}_{p q r}$ is the $(n-3) \times (n-3)$ minor determinant obtained by deleting rows $i,j,k$ and columns $p,q,r$ out of the $n \times n$ symmetric matrix $\phi$, whose entries $\varphi^i{}_{j}$ are
\begin{equation}
\left\{\!\begin{aligned}
& \varphi^{i}{}_{j} = \frac{[ij]}{\langle ij \rangle} \qquad j \neq i \,, \\
& \varphi^{i}{}_{i} = - \sum_{k \neq i} \frac{[ik]\langle kx \rangle \langle ky \rangle}{\langle ik \rangle \langle ix \rangle \langle iy \rangle} \,,
\end{aligned}\right.  
\label{eq:phiij}
\end{equation}
where $x, y$ are arbitrary reference spinors. Up to the sign, the function $\varphi^{i}{}_{i}$ is the gravity soft factor~\cite{Weinberg:1965nx}.

For five gravitons, by choosing the minor $|{\phi}|^{345}_{123}$ the MHV amplitude is
\beq
{\cal M}(1_h^-,2_h^-,3_h^+,4_h^+,5_h^+) =
\angbra{12}^8 \frac{[15][24]\angbra{14}\angbra{25}
- [14][25]\angbra{24}\angbra{15}}{\angbra{12}
    \angbra{23}\angbra{31}\angbra{34}
\angbra{45}\angbra{53}\angbra{14}\angbra{25}\angbra{24}\angbra{15}} \,.
\label{eq:5happ}
\eeq

\section{Six-graviton MHV amplitudes}
\label{app:d}

Using Hodges' formula, \eqnss{eq:amphodges}{eq:phiij}, we display three equivalent representations of the six-graviton MHV amplitude.

Using as minor, $\phi_{123}^{456}$, the six-graviton MHV amplitude is
\begin{equation}
\begin{aligned}
\mathcal{M}(1_h^-,2_h^-,3_h^+,4_h^+,5_h^+,6_h^+)
&=\dfrac{\la 12\ra^7}{\la23\ra\la31\ra\la45\ra\la56\ra\la64\ra} \times\\
&\quad \times\bigg(\dfrac{\lb41\rb\lb52\rb\lb63\rb}{\la41\ra\la52\ra\la63\ra} -
\dfrac{\lb42\rb\lb51\rb\lb63\rb}{\la42\ra\la51\ra\la63\ra}-
\dfrac{\lb41\rb\lb53\rb\lb62\rb}{\la41\ra\la53\ra\la62\ra}+\\
&\hspace{1cm}+
\dfrac{\lb43\rb\lb51\rb\lb62\rb}{\la43\ra\la51\ra\la62\ra}+ \dfrac{\lb42\rb\lb53\rb\lb61\rb}{\la42\ra\la53\ra\la61\ra}-
\dfrac{\lb43\rb\lb52\rb\lb61\rb}{\la43\ra\la52\ra\la61\ra}
\bigg) \,.
\end{aligned}
\label{eq:6gravamp1}
\end{equation}

Using as minor, $\phi_{124}^{356}$, the six-graviton MHV amplitude is
\begin{equation}
\begin{aligned}
\mathcal{M}(1_h^-,2_h^-,3_h^+,4_h^+,5_h^+,6_h^+)
&=\dfrac{\la 12\ra^7}{\la24\ra\la35\ra\la41\ra\la56\ra\la63\ra} \times\\
&\quad \times\bigg(\dfrac{\lb31\rb\lb52\rb\lb64\rb}{\la31\ra\la52\ra\la64\ra} -
\dfrac{\lb32\rb\lb51\rb\lb64\rb}{\la32\ra\la51\ra\la64\ra}-
\dfrac{\lb31\rb\lb54\rb\lb62\rb}{\la31\ra\la54\ra\la62\ra}+\\
&\hspace{1cm}+
\dfrac{\lb34\rb\lb51\rb\lb62\rb}{\la34\ra\la51\ra\la62\ra}+ \dfrac{\lb32\rb\lb54\rb\lb61\rb}{\la32\ra\la54\ra\la61\ra}-
\dfrac{\lb34\rb\lb52\rb\lb61\rb}{\la34\ra\la52\ra\la61\ra}
\bigg) \,.
\end{aligned}
\label{eq:6gravamp2}
\end{equation}

Using as minor, $\phi_{146}^{235}$, the six-graviton MHV amplitude is
\begin{equation}
\begin{aligned}
\mathcal{M}(1_h^-,2_h^-,3_h^+,4_h^+,5_h^+,6_h^+)
&=\dfrac{\la 12\ra^8}{\la14\ra\la23\ra\la35\ra\la46\ra\la52\ra\la61\ra} \times\\
&\quad \times\bigg(\dfrac{\lb21\rb\lb34\rb\lb56\rb}{\la21\ra\la34\ra\la56\ra} -
\dfrac{\lb24\rb\lb31\rb\lb56\rb}{\la24\ra\la31\ra\la56\ra}-
\dfrac{\lb21\rb\lb36\rb\lb54\rb}{\la21\ra\la36\ra\la54\ra}+\\
&\hspace{1cm}+
\dfrac{\lb26\rb\lb31\rb\lb54\rb}{\la26\ra\la31\ra\la54\ra}+ \dfrac{\lb24\rb\lb36\rb\lb51\rb}{\la24\ra\la36\ra\la51\ra}-
\dfrac{\lb26\rb\lb34\rb\lb51\rb}{\la26\ra\la34\ra\la51\ra}
\bigg) \,.
\end{aligned}
\label{eq:6gravamp3}
\end{equation}

\subsection{The MRK limit}
\label{app:dmrk}

In the MRK limit (\ref{eq:mrk}), each of \eqnss{eq:6gravamp1}{eq:6gravamp3} takes the ladder form,
\beqa
\lefteqn{ \lim_{p_3^+\gg p_4^+\gg p_5^+ \gg p_6^+} {\cal M}(1_h^-,2_h^-,3_h^+,4_h^+,5_h^+,6_h^+) } \label{eq:6hmhvreg}\\
&=& - s^2
\left[ \kappa\, C(2_h^-,3_h^+) \right] \frac1{t_1} \left[ \kappa\, V(q_1, 4_h^+, q_2)\right] \frac1{t_2}
\left[ \kappa\, V(q_2, 5_h^+, q_3)\right] \frac1{t_3}
\left[ \kappa\, C(1_h^-,6_h^+) \right] \,, \nn
\eeqa
i.e. explicitly,
\begin{equation}
\begin{aligned}
&\mathcal{M}(1_h^-,2_h^-,3_h^+,4_h^+,5_h^+,6_h^+)\\
&=-\dfrac{s^2}{t_1t_2t_3}\left( \dfrac{p_{6\perp}^*}{p_{6\perp}}\right)^2
\dfrac{q_{1\perp}^*q_{2\perp}}{p_{4\perp}^2}(q_{1\perp}^*q_{2\perp}-q_{2\perp}^*q_{1\perp})
\dfrac{q_{2\perp}^*q_{3\perp}}{p_{5\perp}^2}(q_{2\perp}^*q_{3\perp}-q_{3\perp}^*q_{2\perp}) \,,
\end{aligned}
\label{eq:6hmhvregexpl}
\end{equation}
with $q_1 = -(p_2+p_3)$, $q_2=q_1-p_4$, $q_3= p_1+p_6$, and $t_i= q_i^2\simeq - q_{i\perp} q_{i\perp}^\ast$, with $i = 1, 2, 3$.

\subsection{The NMRK limit}
\label{app:dnmrk}


In the NMRK limit (\ref{eq:nmreq}),
\eqnss{eq:6gravamp1}{eq:6gravamp3} are reduced to
\begin{equation}\label{eq:dam_NMHV}
\begin{aligned}
\mathcal{M}(1_h^-,2_h^-,3_h^+,4_h^+,5_h^+,6_h^+)
&=\\
&\hspace{-4cm}=-\sqrt{\dfrac{p_4^+}{p_5^+}}p_{5\perp}\dfrac{1}{\la45\ra}\bigg\{
-\dfrac{s^2}{t_1t_3}\left( \dfrac{p_{6\perp}^*}{p_{6\perp}}\right)^2
\dfrac{p_{3\perp}^*p_{6\perp}}{p_{4\perp}^2p_{5\perp}^2}
(p_{6\perp}^*p_{5\perp}-p_{5\perp}^*p_{6\perp})
(p_{4\perp}^*p_{3\perp}-p_{3\perp}^*p_{4\perp})\bigg\}
\,+\\
&\hspace{-4cm}\qquad-\sqrt{\dfrac{p_5^+}{p_4^+}}p_{4\perp}\dfrac{1}{\la54\ra}\bigg\{
-\dfrac{s^2}{t_1t_3}\left( \dfrac{p_{6\perp}^*}{p_{6\perp}}\right)^2
\dfrac{p_{3\perp}^*p_{6\perp}}{p_{4\perp}^2p_{5\perp}^2}
(p_{6\perp}^*p_{4\perp}-p_{4\perp}^*p_{6\perp})
(p_{5\perp}^*p_{3\perp}-p_{3\perp}^*p_{5\perp})\bigg\} \,.
\end{aligned}
\end{equation}



\section{Six-graviton NMHV amplitudes}
\label{app:e}

Here we display two equivalent representations of the six-graviton NMHV amplitude. The first is due to Cachazo and Svrcek~\cite{Cachazo:2005ca}
\begin{equation}
    \mathcal{M}(1_h^-,2_h^-,3_h^+,4_h^-,5_h^+,6_h^+)=D_1+D_1^{flip}+D_2+D_3+D_3^{flip}+D_6 \,,
    \label{eq:nmhv6grav1}
\end{equation}
where
\beqa
D_1 &=& \frac{\langle 24\rangle \langle 1|2+4|3]^7 \left([51] \langle 1|2+4|3] \langle 5|4+3|2] + s_{243} [12] [35] \langle 51\rangle\right)}{s_{243} [24] [43]^2 \langle 15\rangle \langle 16\rangle \langle 56\rangle \langle 1|4+3|2] \langle 5|2+4|3] \langle 5|4+3|2] \langle 6|2+4|3] \langle 6|4+3|2]} \nn\\
&+&(1\leftrightarrow2) \,,
\eeqa
\beqa
D_2 &=& -\frac{[16] [35]^7 \langle14\rangle^7 \langle25\rangle}{
    s_{235} [23] [25] \langle16\rangle \langle46\rangle 
    \langle1|2+5|3] \langle4|1+6|2] \langle4|1+6|5] \langle6|2+5|3]} \nn\\
    &+& (1\leftrightarrow2)+(5\leftrightarrow6)
    +\left((1,5)\leftrightarrow (2,6)\right) \,,
\eeqa
\beqa
D_3 &=& \frac{
    [13] [56]^7 \langle14\rangle^8 \bigl([62] \langle24\rangle \langle56\rangle \langle1|4+3|5] + [56] \langle45\rangle \langle62\rangle \langle1|4+3|2]\bigr)
}{
    s_{143} [25] [26] \langle13\rangle \langle43\rangle^2 
    \langle1|4+3|2] \langle1|4+3|5] \langle1|4+3|6] 
    \langle4|1+3|2] \langle4|1+3|5] \langle4|1+3|6]
} \nn\\
&+& (1\leftrightarrow2) \,,
\eeqa
\begin{equation}
D_6=\frac{
    [56] \langle12\rangle \langle4|1+2|3]^8
}{
    s_{123} [13] [21] [23] \langle45\rangle \langle46\rangle \langle56\rangle
    \langle4|5+6|1] \langle4|5+6|2] \langle5|1+2|3] \langle6|1+2|3]
} \,.
\end{equation}
In eq.(\ref{eq:nmhv6grav1}), $D^{flip}$ stands for the flip operation, 
\beq
\langle \rangle \leftrightarrow []\,, \quad i \leftrightarrow 7-i \,. 
\label{eq:flip}
\eeq
Further details on the features of eq.(\ref{eq:nmhv6grav1}) are given in ref.~\cite{Cachazo:2005ca,Trnka:2020dxl}.

The second representation is due to Hodges~\cite{Hodges:2011wm},
\begin{equation}
    \mathcal{M}(1_h^-,2_h^-,3_h^+,4_h^-,5_h^+,6_h^+)=E_1+E_2+E_3+E_3^{flip} \,,
    \label{eq:nmhv6grav2}
\end{equation}
where
\begin{equation}
E_1=-\frac{
    [12]\, [56]^7\, \langle12\rangle^7\, \langle56\rangle
}{
    s_{123}\,[45]\,[46]\,\langle13\rangle\,\langle23\rangle\,
    \langle1|2+3|4]\,\langle2|1+3|4]\,\langle3|1+2|5]\,\langle3|1+2|6]
} \,,
\label{eq:f8}
\end{equation}
\beqa
E_2 &=& \frac{
    [26] \langle15\rangle \langle2|3+6|5]^7
}{
    s_{145} [14] [15] [45] \langle23\rangle \langle26\rangle \langle36\rangle
    \langle3|1+5|4] \langle3|4+5|1] \langle6|1+5|4]
} \nn\\
&+& (1\leftrightarrow2)+(5\leftrightarrow6)
+\left((1,5)\leftrightarrow (2,6)\right) \,,
\eeqa
\beqa
E_3 &=& \frac{
    \langle23\rangle \langle1|2+4|3]^7 
}{
    s_{243} [23] [24] [43] \langle56\rangle 
    \langle1|2+3|4] \langle5|2+3|4] \langle6|2+3|4]
}\left(
        \frac{[15][46]}{\langle15\rangle \langle6|4+3|2]} 
        - 
        \frac{[16][45]}{\langle16\rangle \langle5|4+3|2]}
    \right) \nn\\
    &+& (1\leftrightarrow2) \,,
\eeqa
and\footnote{The coefficient $E_1$ in eq.~(\ref{eq:f8}) differs from the same coefficient in eq.~(83) of ref.~\cite{Hodges:2011wm} by the sign. 
We have checked that with the sign of $E_1$ in eq.~(\ref{eq:f8}) we agree with ref.~\cite{Cachazo:2005ca}, and obtain the correct scaling of the coefficient $b_1$ in eq.~(\ref{eq:bb1}).} the flip operation in eq.~(\ref{eq:nmhv6grav2}) is as in eq.~(\ref{eq:flip}).

\subsection{The NMRK limit}
\label{app:enmrk}

In the NMRK limit, eq.~(\ref{eq:nmhv6grav2}) is reduced to
\beqa
\lefteqn{ \lim_{p_3^+\gg p_4^+\simeq p_5^+ \gg p_6^+} {\cal M}(1_h^-,2_h^-,3_h^+,4_h^-,5_h^+,6_h^+) } \nn\\
&=& - s^2
\left[ \kappa\, C(2_h^-,3_h^+) \right] \frac1{t_1} \left[ \kappa^2\, V(q_1, 4_h^-, 5_h^+, q_3)\right] \frac1{t_3}
\left[ \kappa\, C(1_h^-,6_h^+) \right] \,, \label{eq:6hnmhvnmrkapp}
\eeqa
with the two-graviton CEV of opposite helicities,
\begin{equation}
V(q_1, 4_h^-, 5_h^+, q_3) = - \left(b_1+b_2+b_3+b_4+b_5+b_6\right) \,,
\label{eq:nmhv2gravcev}
\end{equation}
where
\begin{equation}
b_1=\beta\sum_{n=0}^{3}\gamma_n(p_4^+)^n(p_5^+)^{3-n} \,,
\label{eq:bb1}
\end{equation}
with
\begin{equation}
\begin{aligned}
\beta &=w_4^2\dfrac{x_5^3}{x_4 (p^+_4)^2}\dfrac{q_{3\perp}q_{3\perp}^*}{p_4^+p_{5\perp}^*-p_5^+p_{4\perp}^*}\dfrac{q_{1\perp}^*}{(s_{156}+q_{1\perp}q_{2\perp}^*)^2}
\,,\\
\gamma_0 &=(p_{4\perp}^*)^2 (3 p_{4\perp} + 2 p_{5\perp}) (p_{4\perp} + p_{5\perp} + p_{6\perp}) \,,\\
\gamma_1 &= p_{4\perp}^* \bigg\{
p_{4\perp}^* \left( 5 p_{4\perp}^2 + p_{5\perp} (4 p_{5\perp} + 3 p_{6\perp}) + p_{4\perp} (9 p_{5\perp} + 4 p_{6\perp}) \right) \\
& \hspace{1.5cm}-p_{6\perp}^* \left( 3 p_{4\perp}^2 + 3 p_{5\perp} (p_{5\perp} + p_{6\perp}) + p_{4\perp} (6 p_{5\perp} + 4 p_{6\perp}) \right) \\
& \hspace{1.5cm} - p_{5\perp}^* \left( 7 p_{4\perp}^2 + 4 p_{4\perp} (3 p_{5\perp} + 2 p_{6\perp}) + p_{5\perp} (5 p_{5\perp} + 6 p_{6\perp}) \right) \bigg\}\,, \\
\gamma_2 &= (p_{5\perp}^* + p_{6\perp}^*)\Big\{p_{5\perp}^*\big(4p_{4\perp}^2 + p_{5\perp}(3p_{5\perp} + 4p_{6\perp}) + p_{4\perp}(7p_{5\perp} + 5p_{6\perp})\big)-p_{6\perp}^* p_{4\perp}(p_{4\perp} + p_{5\perp})
\Big\} \\
& \hspace{1cm}- p_{4\perp}^*\Big\{
p_{6\perp}^*\big(6p_{4\perp}^2 + 2p_{5\perp}(3p_{5\perp} + 2p_{6\perp}) + p_{4\perp}(12p_{5\perp} + 5p_{6\perp})\big) \\
&\hspace{2.5cm} + p_{5\perp}^*\big(11p_{4\perp}^2 + 10p_{4\perp}(2p_{5\perp} + p_{6\perp}) + p_{5\perp}(9p_{5\perp} + 8p_{6\perp})\big)
\Big\} \,,
 \\
\gamma_3 &=\big(p_{5\perp}^* + p_{6\perp}^*\big) \Big[\, p_{6\perp}^* p_{5\perp} (p_{4\perp} + p_{5\perp})
+\, p_{5\perp}^* (6p_{4\perp} + 5p_{5\perp})(p_{4\perp} + p_{5\perp} + p_{6\perp}) \Big]\,,
\end{aligned}
\end{equation}
\begin{equation}
b_2=\dfrac{q_{3\perp}q_{3\perp}^*}{(p_{4\perp}^*)^2}\dfrac{q_{1\perp}q_{1\perp}^*}{p_{5\perp}^2}\bigg\{(q_{3\perp}^*q_{2\perp}-q_{2\perp}^*q_{3\perp})+p_4^-p_5^+\Big(\dfrac{1}{y_4}-2\dfrac{p_{4\perp}-p_{5\perp}}{p_{4\perp}}\Big)\bigg\} \,,
\end{equation}
\begin{equation}
b_3=\dfrac{q_{3\perp}^2}{(p_{4\perp}^*)^2}
\dfrac{q_{1\perp}q_{1\perp}^*}{p_{5\perp}^2}\dfrac{p_4^+p_5^-}{p_4^-(p_{4\perp}^*p_5^+-p_{5\perp}^*p_4^+)}
\dfrac{(p_{5\perp}^*)^3}{\Big(\dfrac{1}{y_4}+\dfrac{p_{6\perp}}{p_{4\perp}}\Big)}y_5^2 \,,
\end{equation}
\begin{equation}
b_4=\dfrac{(q_{3\perp}^*)^2}{p_{5\perp}^2}
\dfrac{q_{1\perp}^*p_{5\perp}^*(q_{1\perp}-p_{5\perp})^6}{p_{4\perp}^*p_{4\perp}}
\dfrac{x_5}{s_{146}\big(x_5s_{146}+q_{3\perp}^*(x_5 q_{1\perp}-p_{5\perp})\big)\Big(\dfrac{1}{y_4}+\dfrac{p_{6\perp}}{p_{4\perp}}\Big)} \,,
\end{equation}
\beqa
b_5 &=& \dfrac{q_{2\perp}^*q_{3\perp}}{(p_{4\perp}^*)^2}
\dfrac{q_{2\perp}^*q_{1\perp}}{p_{5\perp}^*p_{5\perp}}
\dfrac{(q_{2\perp}^*)^4 p_{4\perp}^2 q_{1\perp}}{(s_{234}+q_{1\perp}q_{2\perp}^*)^2}
\nn\\ &\times&
\dfrac{p_4^+q_{2\perp}^*(q_{3\perp}^*q_{2\perp}-q_{2\perp}^*q_{3\perp})+p_5^+p_{4\perp}^*(q_{3\perp}q_{2\perp}^*-q_{2\perp}q_{3\perp}^*+q_{1\perp}q_{3\perp}^*)}{s_{156}\Big(\dfrac{1}{y_5}+\dfrac{p_{6\perp}^*}{p_{5\perp}^*}\Big)p_4^+p_5^+\Big[p_5^-\Big(\dfrac{1}{2y_5}+\dfrac{p_{6\perp}^*}{p_{5\perp}^*}\Big)+p_4^-\Big(\dfrac{1}{2y_4}+\dfrac{p_{6\perp}}{p_{4\perp}}\Big) \Big]} \,,
\eeqa
\begin{equation}
b_6=\dfrac{q_{1\perp}^*q_{1\perp}}{(p_{4\perp}^*)^2}\dfrac{q_{3\perp}^*}{p_{5\perp}}\dfrac{p_{4\perp}^4}{(p_{4\perp}+p_{5\perp})^2}\dfrac{[45]}{\langle45\rangle}\dfrac{p_5^+}{p_4^+}\dfrac{y_4^2}{s_{156}+q_{3\perp}q_{2\perp}^*}\Big(\dfrac{p_{6\perp}p_{5\perp}^*}{y_5}-\dfrac{p_{6\perp}^*p_{4\perp}}{w_4}\Big) \,,
\end{equation}
where we have defined the momentum fractions,
\begin{equation}
\begin{aligned}
x_i=\dfrac{p_i^+}{p_4^++p_5^+}\,,\qquad
y_i=\dfrac{p_i^-}{p_4^-+p_5^-} \,,\qquad
w_i=\dfrac{p_{i\perp}}{p_{4\perp}+p_{5\perp}} \,.\qquad
\end{aligned}
\end{equation}

\subsection{The MRK limit}
\label{app:emrk}

In the MRK limit, only the $b_2$, $b_3$ e $b_5$ coefficients above survive, and eq.~(\ref{eq:nmhv6grav2}) is reduced to the ladder form,
\beqa
\lefteqn{ \lim_{p_3^+\gg p_4^+\gg p_5^+ \gg p_6^+} {\cal M}(1_h^-,2_h^-,3_h^+,4_h^-,5_h^+,6_h^+) } \label{eq:6hnmhvreg}\\
&=& - s^2
\left[ \kappa\, C(2_h^-,3_h^+) \right] \frac1{t_1} \left[ \kappa\, V(q_1, 4_h^-, q_2)\right] \frac1{t_2}
\left[ \kappa\, V(q_2, 5_h^+, q_3)\right] \frac1{t_3}
\left[ \kappa\, C(1_h^-,6_h^+) \right] \,. \nn
\eeqa

\bibliographystyle{JHEP}
\bibliography{refs.bib}

\end{document}